\documentclass[prl,aps,letterpaper,groupedaddress,twocolumn,superscriptaddress,floatfix]{revtex4-1}

\usepackage[T1]{fontenc}
\usepackage{amstext,amsmath,amssymb,amsthm}
\usepackage[normalem]{ulem}
\usepackage{color}
\usepackage{float} 
\usepackage{times}
\usepackage[table]{xcolor}
\usepackage{enumitem}
\usepackage{graphicx}
\usepackage[breaklinks=true,colorlinks=true,urlcolor=blue,linkcolor=blue,citecolor=red]{hyperref}
\usepackage{siunitx}
\usepackage{bm}
\usepackage{mathtools}
\usepackage{multirow}
\usepackage{comment}

\allowdisplaybreaks[2]

\newcommand{\degree}[1]{\ensuremath{#1^\circ}}
\newcommand{\nuc}[2]{\ensuremath{{}^{#1}\textrm{#2}}}
\newcommand{\Bone}{\ensuremath{\textrm{B}_{1}}}

\newcommand{\ket} [1] {\vert #1 \rangle}
\newcommand{\bra} [1] {\langle #1 \vert}

\makeatletter

\@ifundefined{textcolor}{}
{\definecolor{BLACK}{gray}{0}
 \definecolor{WHITE}{gray}{1}
 \definecolor{RED}{rgb}{1,0,0}
 \definecolor{GREEN}{rgb}{0,.6,0}
 \definecolor{BLUE}{rgb}{0,0,1}
 \definecolor{CYAN}{cmyk}{1,0,0,0}
 \definecolor{MAGENTA}{cmyk}{0,1,0,0}
 \definecolor{YELLOW}{cmyk}{0,0,1,0}}
 \definecolor{light-gray}{gray}{0.90}

\makeatother

\def\id{I}
\def\1{\mat{\id}}
\def\mat#1{\vec{#1}}
\renewcommand{\vec}[1]{\bm{\mathrm{#1}}}

\begin{document} 

\title{Cross-verification of independent quantum devices}

\author{C. Greganti}
\thanks{These authors contributed equally to this work.}
\affiliation{Vienna Center for Quantum Science and Technology (VCQ),
Faculty of Physics, University of Vienna,
Boltzmanngasse 5, Vienna A-1090, Austria}
\affiliation{VitreaLab GmbH, Boltzmanngasse 5, A-1090 Vienna, Austria}
\author{ T. F. Demarie}
\thanks{These authors contributed equally to this work.}
\affiliation{Entropica Labs, 32 Carpenter Street, Singapore 059911}
\affiliation{Centre for Quantum Technologies, National University of Singapore, 3 Science Drive 2, Singapore 117543}
\affiliation{Singapore University of Technology and Design, 8 Somapah Road, Singapore 487372}
\author{M. Ringbauer}
\affiliation{Institut f\"{u}r Experimentalphysik, Universit\"{a}t Innsbruck, Technikerstrasse 25, A-6020 Innsbruck, Austria}
\author{J. A. Jones}
\affiliation{Clarendon Laboratory, Department of Physics, University of Oxford, Parks Road, Oxford OX1 3PU, U.K}
\author{V. Saggio}
\affiliation{Vienna Center for Quantum Science and Technology (VCQ),
Faculty of Physics, University of Vienna,
Boltzmanngasse 5, Vienna A-1090, Austria}
\author{I. A. Calafell}
\affiliation{Vienna Center for Quantum Science and Technology (VCQ),
Faculty of Physics, University of Vienna,
Boltzmanngasse 5, Vienna A-1090, Austria}
\author{L. A. Rozema}
\affiliation{Vienna Center for Quantum Science and Technology (VCQ),
Faculty of Physics, University of Vienna,
Boltzmanngasse 5, Vienna A-1090, Austria}
\author{A. Erhard}
\affiliation{Institut f\"{u}r Experimentalphysik, Universit\"{a}t Innsbruck, Technikerstrasse 25, A-6020 Innsbruck, Austria}
\author{M. Meth}
\affiliation{Institut f\"{u}r Experimentalphysik, Universit\"{a}t Innsbruck, Technikerstrasse 25, A-6020 Innsbruck, Austria}
\author{L. Postler}
\affiliation{Institut f\"{u}r Experimentalphysik, Universit\"{a}t Innsbruck, Technikerstrasse 25, A-6020 Innsbruck, Austria}
\author{R. Stricker}
\affiliation{Institut f\"{u}r Experimentalphysik, Universit\"{a}t Innsbruck, Technikerstrasse 25, A-6020 Innsbruck, Austria}
\author{P. Schindler}
\affiliation{Institut f\"{u}r Experimentalphysik, Universit\"{a}t Innsbruck, Technikerstrasse 25, A-6020 Innsbruck, Austria}
\author{R. Blatt}
\affiliation{Institut f\"{u}r Experimentalphysik, Universit\"{a}t Innsbruck, Technikerstrasse 25, A-6020 Innsbruck, Austria}
\affiliation{Institut f\"{u}r Quantenoptik und Quanteninformation, \"{O}sterreichische Akademie der  Wissenschaften, Otto-Hittmair-Platz 1, A-6020 Innsbruck, Austria}
\author{T. Monz}
\affiliation{Institut f\"{u}r Experimentalphysik, Universit\"{a}t Innsbruck, Technikerstrasse 25, A-6020 Innsbruck, Austria}
\affiliation{Alpine Quantum Technologies GmbH, 6020 Innsbruck, Austria}
\author{P. Walther}
\affiliation{Vienna Center for Quantum Science and Technology (VCQ),
Faculty of Physics, University of Vienna,
Boltzmanngasse 5, Vienna A-1090, Austria}
\author{J. F. Fitzsimons}
\affiliation{Centre for Quantum Technologies, National University of Singapore, 3 Science Drive 2, Singapore 117543}
\affiliation{Singapore University of Technology and Design, 8 Somapah Road, Singapore 487372}
\affiliation{Horizon Quantum Computing, 79 Ayer Rajah Crescent, \#03-01 BASH, Singapore 139955}

\begin{abstract}
Quantum computers are on the brink of surpassing the capabilities of even the most powerful classical computers. This naturally raises the question of how one can trust the results of a quantum computer when they cannot be compared to classical simulation. Here we present a verification technique that exploits the principles of measurement-based quantum computation to link quantum circuits of different input size, depth, and structure. Our approach enables consistency checks of quantum computations within a device, as well as between independent devices. We showcase our protocol by applying it to five state-of-the-art quantum processors, based on four distinct physical architectures: nuclear magnetic resonance, superconducting circuits, trapped ions, and photonics, with up to 6 qubits and 200 distinct circuits.
\end{abstract}

\maketitle
Quantum computers represent a fundamental shift in the way we think about computation. By harnessing quantum interference effects between different possible branches of a computation, quantum processors have the potential to drastically outperform conventional computers for a range of tasks~\cite{Bernstein1997,Grover1997,Simon1997,Shor1999, Nielsen2000,Harrow2017}. Potential applications of quantum computation range from cryptanalysis to the simulation of physical systems and even to machine learning. Extraordinary experimental efforts in recent years have enabled demonstrations of the technology's potential in a growing number of physical systems~\cite{Jones2011,Schindler2013,Barz2014,Devoret2013}. For certain simulation~\cite{Trotzky2012,Braun2015} and sampling~\cite{Arute2019} tasks, these devices are already pushing the limits of classical supercomputers, and it is foreseeable that the next generation of quantum processors will vastly outperform their classical counterparts.

Building such devices, however, remains challenging, with environmental interactions inducing noise that leads to potentially unreliable results for complex computations. This naturally leads to the question whether we can trust the output of a quantum computation, and, more concretely, whether we can certify the output of a computation as correct. The current standard approach is to benchmark the individual quantum gates that make up the computation~\cite{Magesan2011,Erhard2019} to obtain an indication for how well the full system can perform. In practice, however, such an extrapolation is typically unreliable due to effects such as non-Markovian behaviour, spatially and temporally correlated noise, or unmodelled stray interactions~\cite{Wallman2016}. This highlights the need for a complementary technique to establish full system performance.

In order to fill this gap, significant work has been devoted to the development of cryptographically secure verification protocols \cite{fitzsimons2017unconditionally,aharonov2010proceedings,morimae2014verification,hayashi2015verifiable,reichardt2013classical,mckague2016interactive,fitzsimons2018post,coladangelo2017verifier}, with one such technique having been experimentally demonstrated~\cite{barz2013experimental}. However, existing provably secure verification techniques require either quantum communication~\cite{fitzsimons2017unconditionally,aharonov2010proceedings,morimae2014verification,hayashi2015verifiable} or shared entanglement between devices~\cite{reichardt2013classical,mckague2016interactive,fitzsimons2018post,coladangelo2017verifier}, leaving them out of reach for existing quantum processors, which are typically unable to exchange quantum information with one another. A promising technique has recently been proposed to allow verification of a single isolated processor based on computational hardness assumptions~\cite{mahadev2018classical}. However due to large key sizes, implementing this approach would require an extremely sophisticated processor.

\begin{figure}[t]
\begin{center}
\includegraphics[width=0.85\columnwidth]{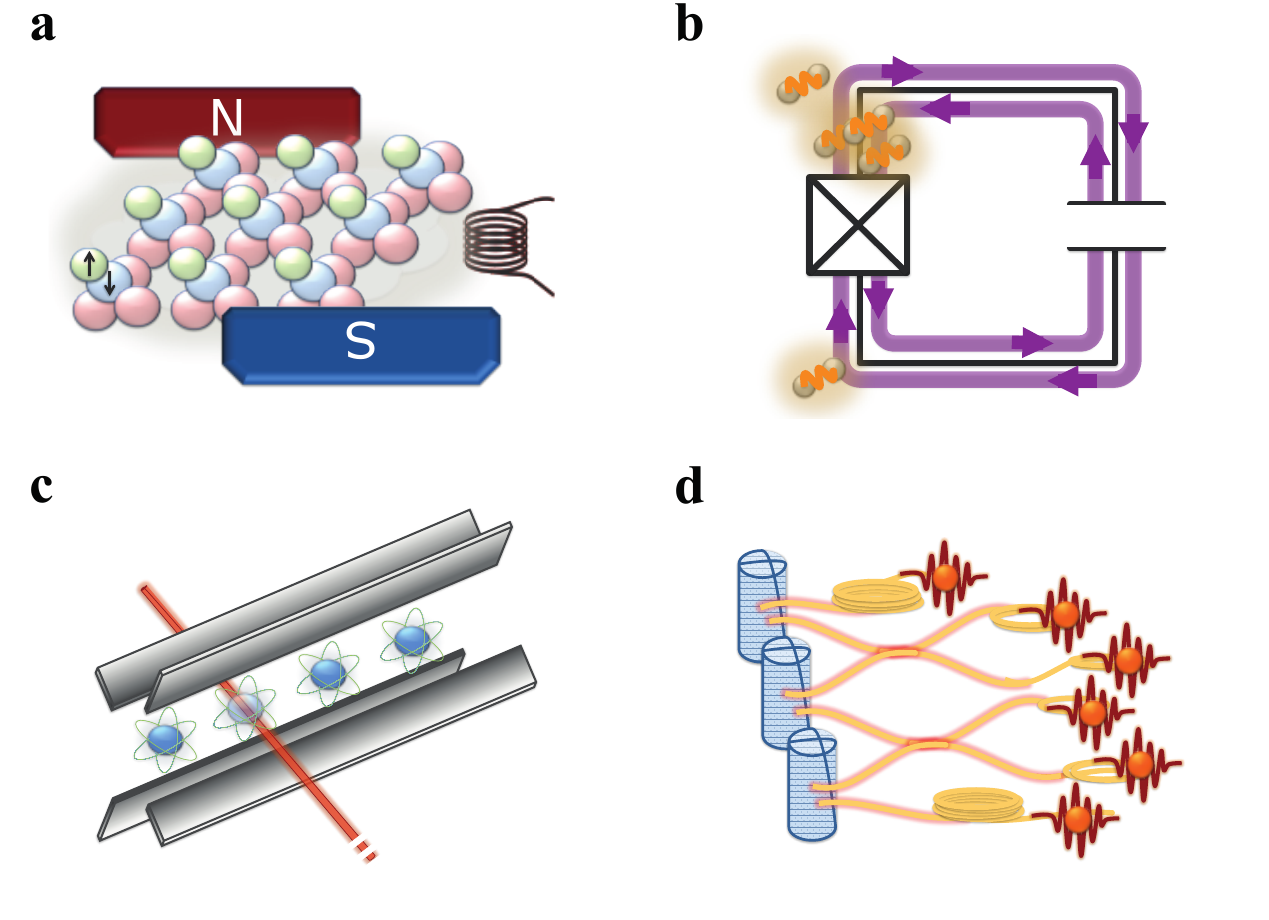} 
\caption{\textbf{A cartoon representation of the quantum processing architectures used.} \textbf{a)} An NMR device at the University of Oxford~\cite{Jones2011}; \textbf{b)} superconducting circuits at IBM~\cite{IBM} and Rigetti Computing~\cite{Rigetti}; \textbf{c)} a trapped-ion quantum processor at the University of Innsbruck~\cite{Schindler2013}; and \textbf{d)} a photonic quantum processor at the University of Vienna~\cite{TunSPS2018}.}
\label{fig:1}
\end{center}
\end{figure}

Here we address the question whether one noisy device can be used to efficiently check another noisy device, without relying on quantum communication or entanglement between devices. We introduce a cross-check procedure that is inherently agnostic to the underlying hardware, sensitive to systematic errors in the implementation, and applicable to any digital quantum computation. In contrast to related work on comparing output states from different quantum devices~\cite{Elben2019}, our approach aims to verify the device rather than a certain output of it. The protocol is built on the framework of measurement-based quantum computation (MBQC)~\cite{Raussendorf2001,BriegelMBQC2009}, which has proven a powerful tool for blind and verifiable computing protocols~\cite{fitzsimons2017private}. By exploiting the intrinsic symmetries of quantum circuits when mapped to an MBQC, our approach allows us to quantitatively compare the outputs of quantum circuits with different size and structure, performed on independent physical devices in any architecture, thus building a high level of trust in the outputs of the device. 

We demonstrate our protocol by running 200 distinct circuits of different width and depth on five state-of-the-art quantum processors, using four primary technologies for digital quantum computation: 1) a nuclear magnetic resonance (NMR) device~\cite{Jones2011} at the University of Oxford, 2) cloud-accessible superconducting systems from IBM~\cite{IBM} and Rigetti~\cite{Rigetti}, 3) a trapped-ion quantum processor~\cite{Schindler2013} at the University of Innsbruck, and 4) a photonic cluster state quantum device~\cite{TunSPS2018} at the University of Vienna, see Fig.~\ref{fig:1}.

\section{From measurement-based quantum computing to correlated sampling problems}
\begin{figure}[b]
\begin{center}
\includegraphics[width=\columnwidth]{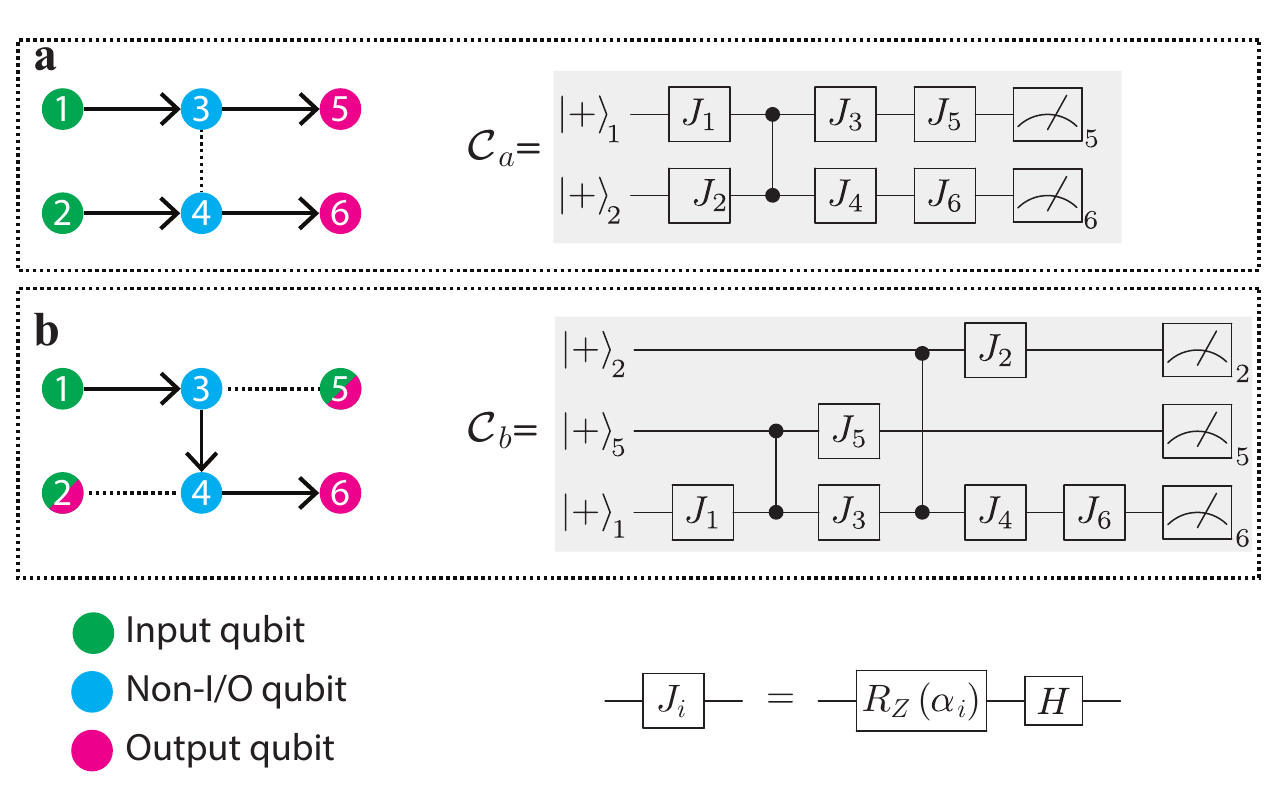}
\caption{\textbf{Schematic representation of equivalent computations in MBQC (left) and in the circuit model (right).} The underlying MBQC is based on a 6-qubit H-shaped graph state with 2 different g-flows, consisting of a) 2 and b) 3 output qubits (other choices are possible). The direction of the flow is indicated by the arrows on the graph, with edges not involved in the flow shown as dashed lines. The qubits are measured according to the order of the labelling numbers. In (a) the input state of the circuit $\mathcal{C}_a$ is $\ket{{++}}_{12}$, associated with qubits $1$ and $2$ of the cluster state, whereas in (b) the input state of the circuit $\mathcal{C}_b$ is $\ket{{+++}}_{251}$, associated with qubits $1$, $2$, and $5$ of the cluster state. The qubit ordering in the circuit was chosen for more intuitive comparison and the detailed procedure for going from the graph and flow to the circuit can be found in the Supplementary Material. Note that both quantum circuits on the right correspond to the same MBQC graph state on the left, albeit with a different flow. The basic gate $\hat{J}(\alpha_i) = \hat{H} \hat{R}_z(\alpha_i)$ (for brevity $\hat{J}_i$ with $i=(1,...,6)$) can be decomposed into a Hadamard gate $\hat{H}$ and a rotation $\hat{R}_z(\alpha)$ around the $\textsc{z}$-axis of the Bloch sphere, see Supplementary Material for details.}
\label{fig:2}
\end{center}
\end{figure}

In order to verify the correctness of a quantum computation we make use of independent runs of several different yet related sampling problems, obtained from a measurement-based implementation of the computation. In contrast to the standard circuit model of quantum computing, where a unitary operation is described by a sequence of gates applied to a reference input, in MBQC a computation is realized as a sequence of single-qubit measurements performed on highly entangled multi-qubit states. These states are also known as \emph{graph states} for their one-to-one correspondence with simple graphs $\mathcal{G} = \{\mathcal{V}, \mathcal{E}\}$. Graph states are represented by a set of vertices $\mathcal{V}$, corresponding to single qubits initialized in the state $\ket{+}=(\ket{0}+\ket{1})/\sqrt{2}$, and a set of edges $\mathcal{E}$, corresponding to pairwise controlled-$\textsc{Z}$ entangling gates applied to the respective vertices, see Fig.~\ref{fig:2} and Supplementary Material for details. The MBQC model is computationally equivalent to the circuit model for appropriate families of graphs~\cite{Raussendorf2001}, even when measurements are restricted to the $\textsc{xy}$-plane of the Bloch sphere~\cite{mantri2016universality}, as considered here.

One way to visualize a quantum computation as an MBQC pattern is to select an (arbitrary) set of input vertices, which represents the initial state of the computation, and an equal-sized output set, which will contain the final state. Fixing these sets determines a unique set of paths connecting each input qubit to an output qubit, giving rise to the notions of \emph{flow}~\cite{Danos2006} and \emph{generalized flow} (g-flow)~\cite{Browne2007}. A deterministic computation then proceeds by sequentially subjecting each non-output qubit along the flow to a projective measurement $P_\alpha = \frac{1}{\sqrt{2}} (\bra{0} \pm e^{-i \alpha} \bra{1})$, at an angle $\alpha$ in the $\textsc{xy}$-plane of the Bloch-sphere, and applying outcome-dependent corrections to the neighbouring qubits. By convention, a measurement outcome of zero requires no correction, whereas a measurement outcome of one requires an update of the measurement angles for subsequent measurements. These flow structures thus specify the possible circuits over a graph by determining the appropriate corrections for non-zero measurement outcomes and their order. The choice of measurement angles $\vec{\alpha} = \{\alpha_1 \ldots \alpha_n\}$ specifies an instance on this circuit.

The key insight that we make use of is that, although MBQC performs a deterministic computation between a specific choice of input and output set, there are always multiple such choices for a given graph. Consequently, there are multiple possible information flows, which is a concept known as \emph{flow ambiguity}~\cite{Mantri2016prx}. These alternate flows give rise to computations with different structure, different number of logical qubits and different flow-dependent corrections, which effectively insert random $\textsc{Z}$-gates into the circuit, see Supplementary Material for details. Thus an MBQC implementing a specific computation can also be seen as providing the outcomes of a random set of other computations, each related to a unique computation in the circuit model. This provides a natural means for testing different devices against each other in a reliable fashion.

We now use this insight to generate MBQC-related sampling problems by converting a given MBQC with angles $\vec{\alpha}$ into the circuit model for different choices of output sets with $n_O$ qubits. To illustrate this, consider the 6-qubit H-shaped graph of Fig.~\ref{fig:2}, where one choice of flow gives rise to the $n_O=2$ qubit circuit $\mathcal{C}_a$ (Fig.~\ref{fig:2}a), while a different choice gives rise to the $n_O=3$ qubit circuit $\mathcal{C}_b$ (Fig.~\ref{fig:2}b). The angles for the elementary single-qubit gates are determined by the measurement angles of the underlying MBQC, and potential additional randomization, see Supplementary Material for details. Our goal is now to relate the outputs of these very different computations.

In the MBQC picture, the measurement outcomes over non-output qubits occur with equal probability of $2^{-(n-n_O)}$~\cite{mhalla2011graph}, while the probabilities for the output qubits ($\text{Pr}$) depends on the choice of measurement angles $\vec \alpha$. Hence, there is no bias, and we can without loss of generality focus on the case of all-zero outcomes for the non-output qubits, where no flow-dependent corrections are necessary. We emphasize, that this choice is arbitrary and merely affects the relation between the measurement angles of the two circuits, but has no effect on the success probability in the circuit picture, where these qubits do not exist. Hence, in the MBQC picture, the probability for obtaining all-zero outcomes using the flow corresponding to $\mathcal{C}_a$ is $2^{-4} \text{Pr}(0,0)_{\mathcal{C}_a}$, where $ \text{Pr}(0,0)_{\mathcal{C}_a}$ is the probability of obtaining zero outcomes when measuring only the $2$ output qubits. Similarly, the probability for all-zero outcomes using the flow of $\mathcal{C}_b$ is $2^{-3} \text{Pr}(0,0,0)_{\mathcal{C}_b}$. Since these two probabilities are obtained from the same graph, they must agree on shared outputs.

Using this fact, we find that the outcome probabilities obtained from the two circuits are related as $\text{Pr}(0,0)_{\mathcal{C}_a}=2 \text{Pr}(0,0,0)_{\mathcal{C}_b}$. Similarly, for the other output combinations we obtain $\text{Pr}(0,1)_{\mathcal{C}_{\text{a}}} = 2 \text{Pr}(0,0,1)_{\mathcal{C}_{\text{b}}}$, $\text{Pr}(1,0)_{\mathcal{C}_{\text{a}}} = 2 \text{Pr}(0,1,0)_{\mathcal{C}_{\text{b}}}$, and $\text{Pr}(1,1)_{\mathcal{C}_{\text{a}}} = 2 \text{Pr}(0,1,1)_{\mathcal{C}_{\text{b}}}$. Note that the case where the remaining qubit of circuit $\mathcal{C}_{\text{b}}$ is in state $1$ can also be used for verification, but is related to a computation with different angles in circuit $\mathcal{C}_{\text{a}}$, see Supplementary Material for more details.

The central observation here is that we can use this technique to establish a connection between the outcome probabilities from two quantum circuits with different width, depth, and structure, but with MBQC-related angles for the single-qubit gates. Implementing these circuits ($\mathcal{C}_a$ and $\mathcal{C}_b$ in the case of the presented experiments) on a single device provides a means for self-verification of the device, while implementing them on different devices provides a pathway to cross-validate the two devices. More generally, all output strings over shared output qubits can be related across circuits, as we describe in detail in the Supplementary Material. Moreover, one can randomize the output strings by adding a random multiple of $\pi$ to the measurement angles for the qubits in the output set. This would allow us to create two distinct, but related circuits such that the probability of obtaining particular (non-zero) strings as outputs is correlated.

\section{Cross-verification}
In order to formally turn this approach into a test of consistency between quantum devices, we now consider two quantum processors, implementing computations derived from the same MBQC, but with different output sets. For output sets of sizes $n_{O_1}$ (processor 1) and $n_{O_2}$ (processor 2), with $n_c$ qubits that are in both output sets, we are left with $n_v = n_{O_1} + n_{O_2} - n_c$ variable bits and thus $2^{n_v}$ different measurement strings $m$ that can be obtained from the two circuits. The output of each quantum processor is a subset of these $n_v$ qubits, allowing for a direct comparison of the devices on this larger space. We denote the vector of probabilities for obtaining the strings $m$ from the quantum circuit $\mathcal{C}_j$ performed on the $j^{\text{th}}$ device (normalized as above) by $\vec{p}_j$. We can now compare the two devices by computing the squared $\ell^2$-distance between the vectors $\vec{p}_j$ (see Supplementary Materials for details)
\begin{equation}
\|\vec{p}_1-\vec{p}_2\|^2 = \vec{p}_1\cdot\vec{p}_1 - 2\vec{p}_1\cdot\vec{p}_2 + \vec{p}_2\cdot\vec{p}_2 .
\label{eq:norm}
\end{equation} 
The terms $\vec{p}_j\cdot\vec{p}_j$ in Eq.~\eqref{eq:norm} are the probabilities of obtaining the same result twice when sampling from the $j^{\text{th}}$ device. These probabilities can be estimated from the minimum number of runs required to obtain a collision among output strings (i.e.\ obtain the same string twice). The MBQC picture then straightforwardly relates the probability for a collision within the $2^{n_v}$ possible strings $m$ to a collision within the $2^{n_{O_j}}$ strings obtained at the circuit output, see Supplementary Materials for details. Consequently estimating $\vec{p}_j\cdot\vec{p}_j$ requires at most $O(2^{n_{O_j}/2})$ runs, independent of the probability distribution, due to a generalization of the birthday paradox~\cite{birthday}. Notably, the values $\vec{p}_j\cdot\vec{p}_j$ also provide a sanity check to detect systems that merely produce uniformly random samples, in which case $\vec{p}_j\cdot\vec{p}_j$ reaches a global minimum. In contrast, deep random quantum circuits are expected to lead to a Porter-Thomas distribution~\cite{Boixo2018}, resulting in values twice as large as in the uniform case. Finally, the term, $\vec{p}_1\cdot\vec{p}_2$ in Eq.~\eqref{eq:norm} can be estimated in a similar as the minimum number of runs to obtain a collision between devices. This is achieved by randomly fixing the value of non-output qubits while sampling over the two devices, which by the same argument as above requires at most $O(2^{(n_{O_1}+n_{O_2}-n_c)/2})$ runs.

Hence, while the exact scaling depends on the problem at hand, in cases where there is a significant number of output qubits in common between the instances ($n_c \sim n_{O_1},n_{O_2}$), or where the output distribution for either computation is far from uniform ($\max \vec{p}_j \gg 2^{-n_{O_j}}$), the quantity $\|\vec{p}_1-\vec{p}_2\|^2$ provides a measure of similarity between the outputs of the devices, which can be estimated with exponentially fewer resources than conventional classical simulation techniques, which typically scale as $2^{n_{O_j}}$~\cite{Arute2019,Pednault2019}.

\section{Experimental Results}
We experimentally performed MBQC-related 2- and 3-qubit circuits with different depth on five independent state-of-the-art quantum processors, covering four of the major quantum computing architectures, see Fig.~\ref{fig:1}. Using these systems, we experimentally implemented sampling instances for the six-qubit H-shaped graph shown in Fig.~\ref{fig:2} by generating $200$ random sets of angles $\{\alpha_i\}_{i=1}^6$, with $\alpha_i\in \{0,\pi/4,\pi/2, 3\pi/4,\pi, 5\pi/4, 3\pi/2, 7\pi/4\}$. We ran 2-qubit circuits of type $\mathcal{C}_a$ on the Oxford and Innsbruck systems, 3-qubit circuits of type $\mathcal{C}_b$ on the Innsbruck, IBM, and Rigetti systems and the 6-qubit H-shaped MBQC on the Vienna system.

After taking into account the relation between the outputs of the different implementations and additional randomization (see Supplementary Material for details), we compute the squared $\ell^2$-distances~\eqref{eq:norm} between pairs of devices implementing different size computations for cross-verification. These values are averaged over $34$ random instances, which is consistent with the results from the full data set, see Supplementary Information. Additionally, we compute the squared $\ell^2$-distance between the 2- and 3-qubit circuits implemented on the Innsbruck device for self-verification. Crucially, in either case we do not compare the output distributions to some ``ideal theory'' (e.g.\ from simulations), which would not be possible for future devices, but rather compare pairs of MBQC-related instances from different devices.

\begin{figure}[t]
\begin{center}
\includegraphics[width=0.9\columnwidth]{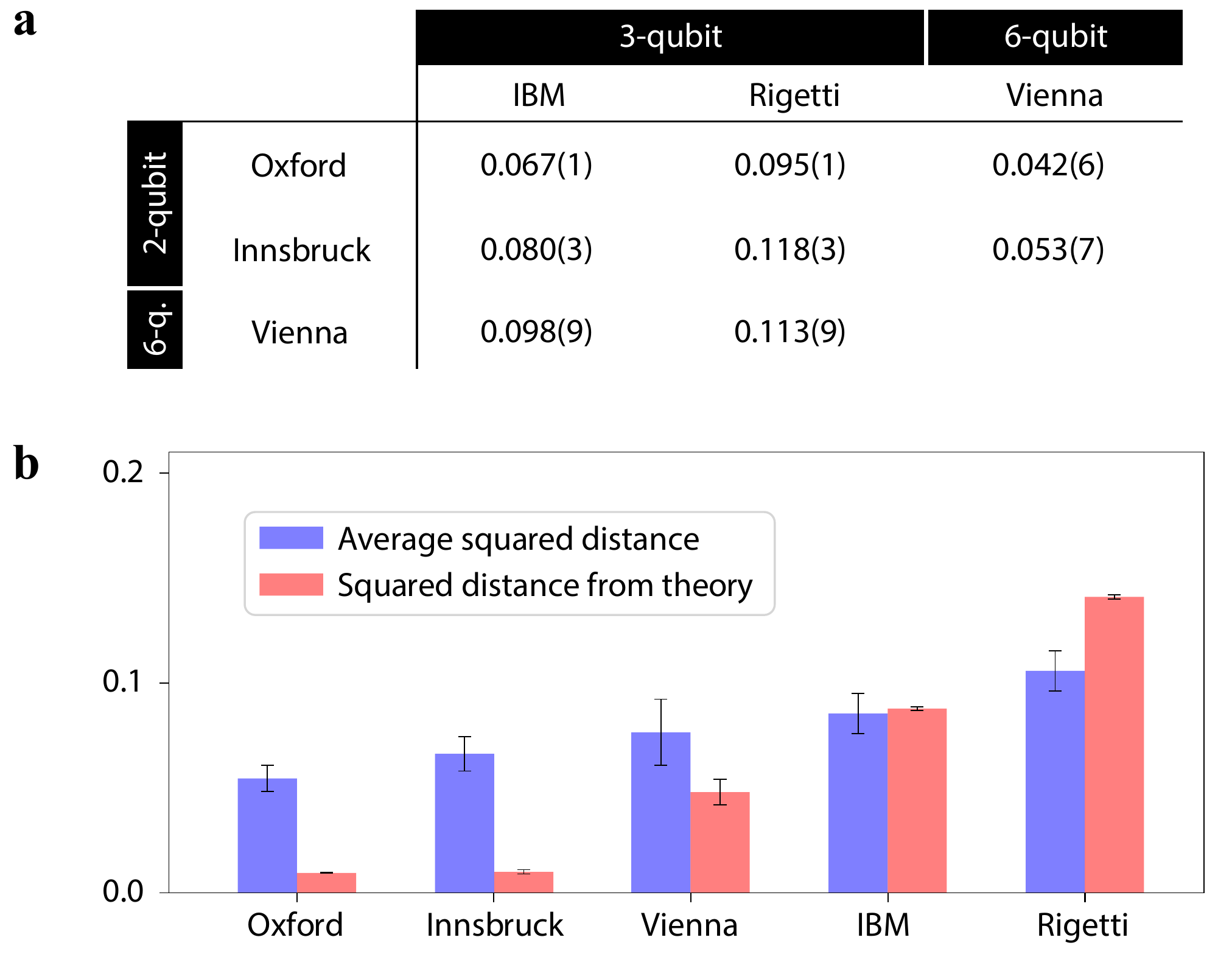}
\caption{\textbf{Experimental comparison of quantum devices.} \textbf{a)} Experimental squared $\ell^2$-distances between pairs of independent quantum devices, averaged over 34 instances per device. Only computations using different numbers of physical qubits are compared, so that the implementations represent fundamentally different sampling problems. Uncertainties in parenthesis correspond to one standard deviation of statistical noise.
\textbf{b)} Squared $\ell^2$-distances for each quantum device averaged over all comparisons to the other devices (blue). Also shown are the squared $\ell^2$-distances of each device against the (non-scalable) theory prediction (red). These two quantities are not expected to coincide. However, they show some qualitative agreement, in that arranging devices according to either metric yields the same order in our experiments. Averages are taken over squared $\ell^2$-distance between each device and all other devices, not just the ones in a), in order to avoid bias.}
\label{fig:results}
\end{center}
\end{figure}

\begin{figure*}[t!]
\begin{center}
\includegraphics[width=0.85\linewidth]{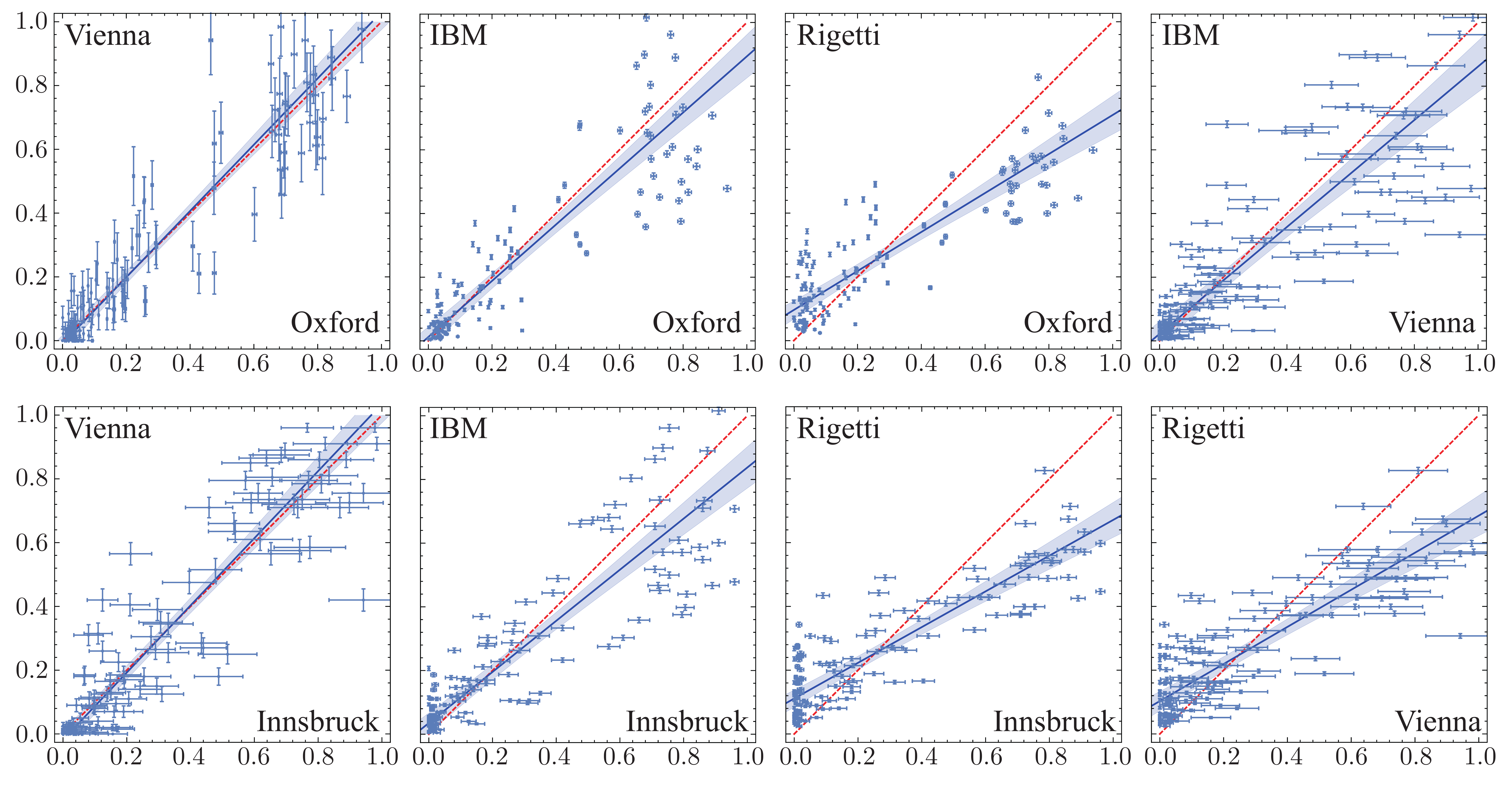} \caption{\textbf{Experimental comparison between outputs of computations for pairs of different devices.} Data points represent the (rescaled) outcome probabilities of one device (horizontal axis, bottom label) versus another (vertical axis, top label), for MBQC-related instances. This is done for all pairs of independent devices implementing different-size circuits  and based on different physical systems. In the ideal case, all points should, up to unavoidable shot noise, lie on the diagonal. For each dataset containing 136 samples (34 circuits and 4 outcome combinations each), we performed linear total least-squares regression (solid blue line) to quantify the deviation from the ideal correlation (red dashed line), yielding regression slopes with 1-sigma uncertainties of (top, left to right) $1.04(2)$, $0.88(3)$, $0.61(2)$, $0.84(3)$, (bottom, left to right) $1.05(2)$, $0.80(3)$, $0.56(2)$, and $0.58(2)$, respectively. Experimental error bars correspond to 1-sigma statistical uncertainty associated to the data points, and the blue shaded bands represent 3-sigma mean prediction intervals for the regression.}
\label{fig:CompPairwise}
\end{center}
\end{figure*}

The key results of these comparisons are the estimated squared $\ell^2$-distance for each pair of devices performing computations of different size and depth, shown in Fig.~\ref{fig:results}a. A value close to 0, for example between Vienna and Oxford or Vienna and Innsbruck, indicates agreement between the devices, whereas any systematic error or statistical noise will lead to a larger value. For example, comparing an ideal 2-qubit circuit to a fully depolarized circuit for the instances considered in our experiments would return a value of $\sim 0.428$. In the limit of large deep random circuits, the squared $\ell^2$-distance between an ideal and fully depolarized circuit converges to $2^{-n_v}$. Of course, the noise in real experiments is much more complicated and the exact dependence of the squared $\ell^2$-distance on such physical noise models remains an interesting question for future research.

When more than two devices are used, one can furthermore make a prediction about the performance of the individual devices by averaging the squared $\ell^2$-distance with all other devices, see Fig.~\ref{fig:results}b. For the small number of qubits involved here, we can still compute the ideal output distribution of each circuit classically. Computing the $\ell^2$-distance with this ideal theory prediction quantifies the true accuracy of each device. Although these values will not be available for future devices, we find that they qualitatively agree with the average squared $\ell^2$-distance over all experimental comparisons. This indicates that the latter provide a good estimate for the true system performance, while the values from individual comparisons accurately capture the relative performance of the devices, thus enabling verification of the underlying quantum processors.

Besides cross-verification between dissimilar quantum devices, our method also provides an intriguing pathway towards self-verification of a single device. Using the Innsbruck trapped-ion system, we implemented MBQC-related instances of the 2- and 3-qubit circuits $\mathcal{C}_a$ and $\mathcal{C}_b$ of Fig.~\ref{fig:2}, and estimate ${\|\vec{p}_{\mathcal{C}_a}-\vec{p}_{\mathcal{C}_b}\|^2 = 0.033(1)}$. This result indicates very good (relative to the results in Fig.~\ref{fig:results}a) agreement between the two circuits, which is confirmed by direct comparison to theory. Notably, by virtue of sampling from multiple instances of vastly different circuits, even systematic errors would be detected, as they manifest very differently in the two circuits. This demonstrates that our method can be used for independent verification of a single quantum processor.

\section{Verification for NISQ devices}
In order to gain some insight into the measured $\ell^2$-distances, correlation plots between the MBQC-related outcomes on all pairs of devices computing different size and depth circuits are shown in Fig.~\ref{fig:CompPairwise}. We emphasize that it is not scalable to produce such plots for larger computations, as they require sampling the full output distribution for multiple computations. However, for near-term intermediate scale (NISQ) devices, this remains feasible and provides useful additional information to aid the interpretation of the $\ell^2$-distances used for scalable cross-verification in Fig.~\ref{fig:results}.

\begin{figure*}[t!]
\begin{center}
\includegraphics[width=0.95\linewidth]{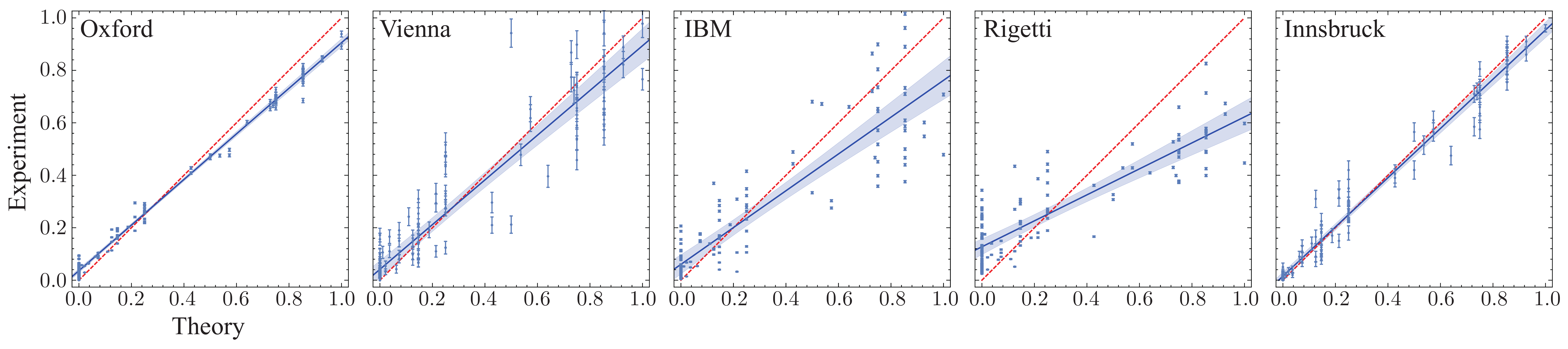}
\caption{\textbf{Comparison between experimental outcome probabilities and theoretical expected values per single device.} From left to right: Oxford, Vienna, IBM, Rigetti, Innsbruck. As in Fig.~\ref{fig:CompPairwise}, data points represent the (rescaled) outcome probabilities obtained from the respective device (Experiment), against the corresponding theory value (Theory) obtained from direct circuit simulation. For each data set we performed linear least-squares regression (blue line) to quantify the deviation from the ideal correlation (red dashed line), yielding regression slopes with 1-sigma uncertainties of $0.869(6)$, $0.85(3)$, $0.70(3)$, $0.50(2)$, and $0.94(1)$, respectively. Experimental error bars correspond to 1-sigma statistical uncertainty and the blue shaded bands represent 3-sigma mean prediction intervals for the regression.}
\label{fig:Theory}
\end{center}
\end{figure*}

\begin{figure}[t!]
\begin{center}
\includegraphics[width=0.65\linewidth]{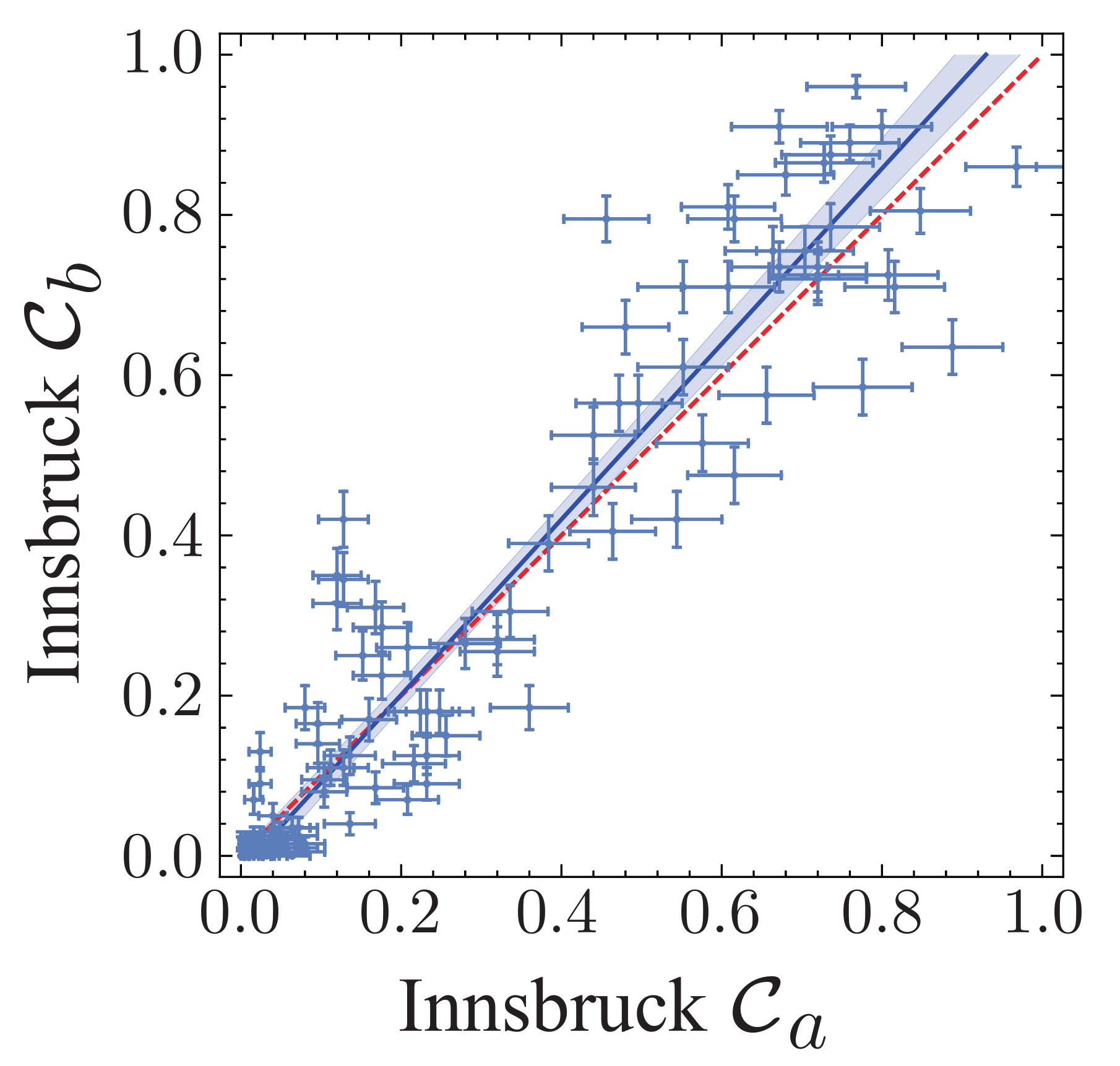}
\caption{\textbf{Comparison between experimental outcome probabilities for the circuits $\mathcal{C}_a$ and $\mathcal{C}_b$ on the Innsbruck device.} Linear total least-squares regression (solid blue line) yields a regression slope of $1.09(2)$ compared to the ideal value of 1 (dashed red line). Experimental error bars correspond to 1-sigma statistical uncertainty and the blue shaded bands represent 3-sigma mean prediction intervals for the regression.}
\label{fig:CompInternal}
\end{center}
\end{figure}

While these correlation plots only provide a crude indication of the strength and direction of the correlations between the devices, there are some notable features. In the ideal case, one expects the two MBQC-related circuits to produce identical outcome probabilities, resulting in clustering around the $45^\circ$ line. On the other hand, depolarizing noise affecting the device on the vertical (horizontal) axis, would result in a mean correlation with slope smaller (larger) than 1. Large scattering around the mean further indicates higher levels of noise in one or both devices, since each point corresponds to a different output or instance of the computation. In addition, Fig.~\ref{fig:Theory} shows correlation plots of the experimental outcomes per individual device against the respective ideal theory prediction obtained via direct simulation of the corresponding circuits.   Fig.~\ref{fig:CompInternal} shows correlation plots  of experimental outcomes  between the 2- and 3-qubit circuits, respectively $\mathcal{C}_a$ and $\mathcal{C}_b$,  performed on the Innsbruck device for self-verification.

\section{Discussion}
As quantum processors start to surpass their classical counterparts, verification by direct comparison to theory will no longer be an option. The technique we present here provides a feasible alternative by validating quantum devices against each other. While not providing a complete toolkit for characterization of individual quantum processors, our method takes a crucial step away from the dependence on classical methods. By sampling from different physical devices implementing circuits that differ in the number of qubits, depth and structure, our method is robust against systematic, as well as statistical errors. By implementing these dissimilar circuits on a single device, our method also provides an avenue towards internal self-verification of single quantum devices.

A particularly intriguing feature of our approach is the way in which it allows us to compare devices using radically different implementations. Recently, a detailed comparison of a trapped ion system and a superconducting processor highlighted the advantages of each system for certain, identical problems~\cite{Linke2017}, concluding that each processor was well suited to different tasks. In this work we overcome the heterogeneity of quantum physical systems, introducing a verification model which links computational circuits with different sizes and depths, and consequently can be run on the many types of quantum computer. The building block of our cross-check scheme is represented by measurement-based quantum computation, which has been proven to be already essential for quantum computation security~\cite{Barz2012}, quantum error correction~\cite{Lanyon2013}, as well as quantum simulation~\cite{Cutcheon2018}. This will prove useful in providing consistent benchmarks across the increasingly diverse range of quantum processors.

\footnotesize
\bibliographystyle{apsrev}
\bibliography{biblio}

\begin{thebibliography}{63}
\expandafter\ifx\csname natexlab\endcsname\relax\def\natexlab#1{#1}\fi
\expandafter\ifx\csname bibnamefont\endcsname\relax
  \def\bibnamefont#1{#1}\fi
\expandafter\ifx\csname bibfnamefont\endcsname\relax
  \def\bibfnamefont#1{#1}\fi
\expandafter\ifx\csname citenamefont\endcsname\relax
  \def\citenamefont#1{#1}\fi
\expandafter\ifx\csname url\endcsname\relax
  \def\url#1{\texttt{#1}}\fi
\expandafter\ifx\csname urlprefix\endcsname\relax\def\urlprefix{URL }\fi
\providecommand{\bibinfo}[2]{#2}
\providecommand{\eprint}[2][]{\url{#2}}

\bibitem[{\citenamefont{Bernstein and Vazirani}(1997)}]{Bernstein1997}
\bibinfo{author}{\bibfnamefont{E.}~\bibnamefont{Bernstein}} \bibnamefont{and}
  \bibinfo{author}{\bibfnamefont{U.}~\bibnamefont{Vazirani}},
  \bibinfo{journal}{SIAM J. Comput.} \textbf{\bibinfo{volume}{26}},
  \bibinfo{pages}{1411} (\bibinfo{year}{1997}).

\bibitem[{\citenamefont{Grover}(1997)}]{Grover1997}
\bibinfo{author}{\bibfnamefont{L.~K.} \bibnamefont{Grover}},
  \bibinfo{journal}{Phys. Rev. Lett.} \textbf{\bibinfo{volume}{79}},
  \bibinfo{pages}{325} (\bibinfo{year}{1997}).

\bibitem[{\citenamefont{Simon}(1997)}]{Simon1997}
\bibinfo{author}{\bibfnamefont{D.~R.} \bibnamefont{Simon}},
  \bibinfo{journal}{SIAM J. Comp.} \textbf{\bibinfo{volume}{26}},
  \bibinfo{pages}{1474} (\bibinfo{year}{1997}).

\bibitem[{\citenamefont{Shor}(1999)}]{Shor1999}
\bibinfo{author}{\bibfnamefont{P.~W.} \bibnamefont{Shor}},
  \bibinfo{journal}{SIAM Rev.} \textbf{\bibinfo{volume}{41}},
  \bibinfo{pages}{303} (\bibinfo{year}{1999}).

\bibitem[{\citenamefont{Nielsen and Chuang}(2000)}]{Nielsen2000}
\bibinfo{author}{\bibfnamefont{M.}~\bibnamefont{Nielsen}} \bibnamefont{and}
  \bibinfo{author}{\bibfnamefont{I.}~\bibnamefont{Chuang}},
  \emph{\bibinfo{title}{Quantum Computation and Quantum Information}}
  (\bibinfo{publisher}{Cambridge University Press}, \bibinfo{year}{2000}).

\bibitem[{\citenamefont{Harrow and Montanaro}(2017)}]{Harrow2017}
\bibinfo{author}{\bibfnamefont{A.~W.} \bibnamefont{Harrow}} \bibnamefont{and}
  \bibinfo{author}{\bibfnamefont{A.}~\bibnamefont{Montanaro}},
  \bibinfo{journal}{Nature} \textbf{\bibinfo{volume}{549}},
  \bibinfo{pages}{203} (\bibinfo{year}{2017}).

\bibitem[{\citenamefont{Jones}(2011)}]{Jones2011}
\bibinfo{author}{\bibfnamefont{J.~A.} \bibnamefont{Jones}},
  \bibinfo{journal}{Prog. Nucl. Magn. Reson. Spectrosc.} pp.
  \bibinfo{pages}{91--120} (\bibinfo{year}{2011}).

\bibitem[{\citenamefont{Schindler et~al.}(2013)\citenamefont{Schindler, Nigg,
  Monz, Barreiro, Martinez, Wang, Quint, Brandl, Nebendahl, Roos
  et~al.}}]{Schindler2013}
\bibinfo{author}{\bibfnamefont{P.}~\bibnamefont{Schindler}},
  \bibinfo{author}{\bibfnamefont{D.}~\bibnamefont{Nigg}},
  \bibinfo{author}{\bibfnamefont{T.}~\bibnamefont{Monz}},
  \bibinfo{author}{\bibfnamefont{J.~T.} \bibnamefont{Barreiro}},
  \bibinfo{author}{\bibfnamefont{E.}~\bibnamefont{Martinez}},
  \bibinfo{author}{\bibfnamefont{S.~X.} \bibnamefont{Wang}},
  \bibinfo{author}{\bibfnamefont{S.}~\bibnamefont{Quint}},
  \bibinfo{author}{\bibfnamefont{M.~F.} \bibnamefont{Brandl}},
  \bibinfo{author}{\bibfnamefont{V.}~\bibnamefont{Nebendahl}},
  \bibinfo{author}{\bibfnamefont{C.~F.} \bibnamefont{Roos}},
  \bibnamefont{et~al.}, \bibinfo{journal}{New J. Phys.}
  \textbf{\bibinfo{volume}{15}}, \bibinfo{pages}{123012}
  (\bibinfo{year}{2013}).

\bibitem[{\citenamefont{Barz et~al.}(2015)\citenamefont{Barz, Kassal,
  Ringbauer, Lipp, Daki{\'{c}}, Aspuru-Guzik, and Walther}}]{Barz2014}
\bibinfo{author}{\bibfnamefont{S.}~\bibnamefont{Barz}},
  \bibinfo{author}{\bibfnamefont{I.}~\bibnamefont{Kassal}},
  \bibinfo{author}{\bibfnamefont{M.}~\bibnamefont{Ringbauer}},
  \bibinfo{author}{\bibfnamefont{Y.~O.} \bibnamefont{Lipp}},
  \bibinfo{author}{\bibfnamefont{B.}~\bibnamefont{Daki{\'{c}}}},
  \bibinfo{author}{\bibfnamefont{A.}~\bibnamefont{Aspuru-Guzik}},
  \bibnamefont{and} \bibinfo{author}{\bibfnamefont{P.}~\bibnamefont{Walther}},
  \bibinfo{journal}{Sci. Rep.} \textbf{\bibinfo{volume}{4}},
  \bibinfo{pages}{6115} (\bibinfo{year}{2015}).

\bibitem[{\citenamefont{Devoret and Schoelkopf}(2013)}]{Devoret2013}
\bibinfo{author}{\bibfnamefont{M.~H.} \bibnamefont{Devoret}} \bibnamefont{and}
  \bibinfo{author}{\bibfnamefont{R.~J.} \bibnamefont{Schoelkopf}},
  \bibinfo{journal}{Science} \textbf{\bibinfo{volume}{339}},
  \bibinfo{pages}{1169} (\bibinfo{year}{2013}).

\bibitem[{\citenamefont{Trotzky et~al.}(2012)\citenamefont{Trotzky, Chen,
  Flesch, McCulloch, Schollw{\"{o}}ck, Eisert, and Bloch}}]{Trotzky2012}
\bibinfo{author}{\bibfnamefont{S.}~\bibnamefont{Trotzky}},
  \bibinfo{author}{\bibfnamefont{Y.-A.} \bibnamefont{Chen}},
  \bibinfo{author}{\bibfnamefont{A.}~\bibnamefont{Flesch}},
  \bibinfo{author}{\bibfnamefont{I.~P.} \bibnamefont{McCulloch}},
  \bibinfo{author}{\bibfnamefont{U.}~\bibnamefont{Schollw{\"{o}}ck}},
  \bibinfo{author}{\bibfnamefont{J.}~\bibnamefont{Eisert}}, \bibnamefont{and}
  \bibinfo{author}{\bibfnamefont{I.}~\bibnamefont{Bloch}},
  \bibinfo{journal}{Nat. Phys.} \textbf{\bibinfo{volume}{8}},
  \bibinfo{pages}{325} (\bibinfo{year}{2012}).

\bibitem[{\citenamefont{Braun et~al.}(2015)\citenamefont{Braun, Friesdorf,
  Hodgman, Schreiber, Ronzheimer, Riera, del Rey, Bloch, Eisert, and
  Schneider}}]{Braun2015}
\bibinfo{author}{\bibfnamefont{S.}~\bibnamefont{Braun}},
  \bibinfo{author}{\bibfnamefont{M.}~\bibnamefont{Friesdorf}},
  \bibinfo{author}{\bibfnamefont{S.~S.} \bibnamefont{Hodgman}},
  \bibinfo{author}{\bibfnamefont{M.}~\bibnamefont{Schreiber}},
  \bibinfo{author}{\bibfnamefont{J.~P.} \bibnamefont{Ronzheimer}},
  \bibinfo{author}{\bibfnamefont{A.}~\bibnamefont{Riera}},
  \bibinfo{author}{\bibfnamefont{M.}~\bibnamefont{del Rey}},
  \bibinfo{author}{\bibfnamefont{I.}~\bibnamefont{Bloch}},
  \bibinfo{author}{\bibfnamefont{J.}~\bibnamefont{Eisert}}, \bibnamefont{and}
  \bibinfo{author}{\bibfnamefont{U.}~\bibnamefont{Schneider}},
  \bibinfo{journal}{Proc. Natl. Acad. Sci.} \textbf{\bibinfo{volume}{112}},
  \bibinfo{pages}{3641} (\bibinfo{year}{2015}).

\bibitem[{\citenamefont{Arute et~al.}(2019)\citenamefont{Arute, Arya, Babbush,
  Bacon, Bardin, Barends, Biswas, Boixo, Brandao, Buell et~al.}}]{Arute2019}
\bibinfo{author}{\bibfnamefont{F.}~\bibnamefont{Arute}},
  \bibinfo{author}{\bibfnamefont{K.}~\bibnamefont{Arya}},
  \bibinfo{author}{\bibfnamefont{R.}~\bibnamefont{Babbush}},
  \bibinfo{author}{\bibfnamefont{D.}~\bibnamefont{Bacon}},
  \bibinfo{author}{\bibfnamefont{J.~C.} \bibnamefont{Bardin}},
  \bibinfo{author}{\bibfnamefont{R.}~\bibnamefont{Barends}},
  \bibinfo{author}{\bibfnamefont{R.}~\bibnamefont{Biswas}},
  \bibinfo{author}{\bibfnamefont{S.}~\bibnamefont{Boixo}},
  \bibinfo{author}{\bibfnamefont{F.~G. S.~L.} \bibnamefont{Brandao}},
  \bibinfo{author}{\bibfnamefont{D.~A.} \bibnamefont{Buell}},
  \bibnamefont{et~al.}, \bibinfo{journal}{Nature}
  \textbf{\bibinfo{volume}{574}}, \bibinfo{pages}{505} (\bibinfo{year}{2019}).

\bibitem[{\citenamefont{Magesan et~al.}(2011)\citenamefont{Magesan, Gambetta,
  and Emerson}}]{Magesan2011}
\bibinfo{author}{\bibfnamefont{E.}~\bibnamefont{Magesan}},
  \bibinfo{author}{\bibfnamefont{J.~M.} \bibnamefont{Gambetta}},
  \bibnamefont{and} \bibinfo{author}{\bibfnamefont{J.}~\bibnamefont{Emerson}},
  \bibinfo{journal}{Phys. Rev. Lett.} \textbf{\bibinfo{volume}{106}},
  \bibinfo{pages}{180504} (\bibinfo{year}{2011}).

\bibitem[{\citenamefont{Erhard et~al.}(2019)\citenamefont{Erhard, Wallman,
  Postler, Meth, Stricker, Martinez, Schindler, Monz, Emerson, and
  Blatt}}]{Erhard2019}
\bibinfo{author}{\bibfnamefont{A.}~\bibnamefont{Erhard}},
  \bibinfo{author}{\bibfnamefont{J.~J.} \bibnamefont{Wallman}},
  \bibinfo{author}{\bibfnamefont{L.}~\bibnamefont{Postler}},
  \bibinfo{author}{\bibfnamefont{M.}~\bibnamefont{Meth}},
  \bibinfo{author}{\bibfnamefont{R.}~\bibnamefont{Stricker}},
  \bibinfo{author}{\bibfnamefont{E.~A.} \bibnamefont{Martinez}},
  \bibinfo{author}{\bibfnamefont{P.}~\bibnamefont{Schindler}},
  \bibinfo{author}{\bibfnamefont{T.}~\bibnamefont{Monz}},
  \bibinfo{author}{\bibfnamefont{J.}~\bibnamefont{Emerson}}, \bibnamefont{and}
  \bibinfo{author}{\bibfnamefont{R.}~\bibnamefont{Blatt}},
  \bibinfo{journal}{Nat. Commun.} \textbf{\bibinfo{volume}{10}},
  \bibinfo{pages}{5347} (\bibinfo{year}{2019}).

\bibitem[{\citenamefont{Wallman and Emerson}(2016)}]{Wallman2016}
\bibinfo{author}{\bibfnamefont{J.~J.} \bibnamefont{Wallman}} \bibnamefont{and}
  \bibinfo{author}{\bibfnamefont{J.}~\bibnamefont{Emerson}},
  \bibinfo{journal}{Phys. Rev. A} \textbf{\bibinfo{volume}{94}},
  \bibinfo{pages}{052325} (\bibinfo{year}{2016}).

\bibitem[{\citenamefont{Fitzsimons and
  Kashefi}(2017)}]{fitzsimons2017unconditionally}
\bibinfo{author}{\bibfnamefont{J.~F.} \bibnamefont{Fitzsimons}}
  \bibnamefont{and} \bibinfo{author}{\bibfnamefont{E.}~\bibnamefont{Kashefi}},
  \bibinfo{journal}{Physical Review A} \textbf{\bibinfo{volume}{96}},
  \bibinfo{pages}{012303} (\bibinfo{year}{2017}).

\bibitem[{\citenamefont{Aharonov et~al.}(2010)\citenamefont{Aharonov, Ben-Or,
  and Eban}}]{aharonov2010proceedings}
\bibinfo{author}{\bibfnamefont{D.}~\bibnamefont{Aharonov}},
  \bibinfo{author}{\bibfnamefont{M.}~\bibnamefont{Ben-Or}}, \bibnamefont{and}
  \bibinfo{author}{\bibfnamefont{E.}~\bibnamefont{Eban}},
  \bibinfo{journal}{Proceedings of Innovations in Computer Science} pp.
  \bibinfo{pages}{453--469} (\bibinfo{year}{2010}).

\bibitem[{\citenamefont{Morimae}(2014)}]{morimae2014verification}
\bibinfo{author}{\bibfnamefont{T.}~\bibnamefont{Morimae}},
  \bibinfo{journal}{Physical Review A} \textbf{\bibinfo{volume}{89}},
  \bibinfo{pages}{060302} (\bibinfo{year}{2014}).

\bibitem[{\citenamefont{Hayashi and Morimae}(2015)}]{hayashi2015verifiable}
\bibinfo{author}{\bibfnamefont{M.}~\bibnamefont{Hayashi}} \bibnamefont{and}
  \bibinfo{author}{\bibfnamefont{T.}~\bibnamefont{Morimae}},
  \bibinfo{journal}{Physical review letters} \textbf{\bibinfo{volume}{115}},
  \bibinfo{pages}{220502} (\bibinfo{year}{2015}).

\bibitem[{\citenamefont{Reichardt et~al.}(2013)\citenamefont{Reichardt, Unger,
  and Vazirani}}]{reichardt2013classical}
\bibinfo{author}{\bibfnamefont{B.~W.} \bibnamefont{Reichardt}},
  \bibinfo{author}{\bibfnamefont{F.}~\bibnamefont{Unger}}, \bibnamefont{and}
  \bibinfo{author}{\bibfnamefont{U.}~\bibnamefont{Vazirani}},
  \bibinfo{journal}{Nature} \textbf{\bibinfo{volume}{496}},
  \bibinfo{pages}{456} (\bibinfo{year}{2013}).

\bibitem[{\citenamefont{McKague}(2016)}]{mckague2016interactive}
\bibinfo{author}{\bibfnamefont{M.}~\bibnamefont{McKague}},
  \bibinfo{journal}{Theory of Computing} \textbf{\bibinfo{volume}{12}},
  \bibinfo{pages}{1} (\bibinfo{year}{2016}).

\bibitem[{\citenamefont{Fitzsimons et~al.}(2018)\citenamefont{Fitzsimons,
  Hajdu{\v{s}}ek, and Morimae}}]{fitzsimons2018post}
\bibinfo{author}{\bibfnamefont{J.~F.} \bibnamefont{Fitzsimons}},
  \bibinfo{author}{\bibfnamefont{M.}~\bibnamefont{Hajdu{\v{s}}ek}},
  \bibnamefont{and} \bibinfo{author}{\bibfnamefont{T.}~\bibnamefont{Morimae}},
  \bibinfo{journal}{Physical review letters} \textbf{\bibinfo{volume}{120}},
  \bibinfo{pages}{040501} (\bibinfo{year}{2018}).

\bibitem[{\citenamefont{Coladangelo et~al.}(2019)\citenamefont{Coladangelo,
  Grilo, Jeffery, and Vidick}}]{coladangelo2017verifier}
\bibinfo{author}{\bibfnamefont{A.}~\bibnamefont{Coladangelo}},
  \bibinfo{author}{\bibfnamefont{A.}~\bibnamefont{Grilo}},
  \bibinfo{author}{\bibfnamefont{S.}~\bibnamefont{Jeffery}}, \bibnamefont{and}
  \bibinfo{author}{\bibfnamefont{T.}~\bibnamefont{Vidick}},
  \bibinfo{journal}{Annual International Conference on the Theory and
  Applications of Cryptographic Techniques} pp. \bibinfo{pages}{247--277}
  (\bibinfo{year}{2019}).

\bibitem[{\citenamefont{Barz et~al.}(2013)\citenamefont{Barz, Fitzsimons,
  Kashefi, and Walther}}]{barz2013experimental}
\bibinfo{author}{\bibfnamefont{S.}~\bibnamefont{Barz}},
  \bibinfo{author}{\bibfnamefont{J.~F.} \bibnamefont{Fitzsimons}},
  \bibinfo{author}{\bibfnamefont{E.}~\bibnamefont{Kashefi}}, \bibnamefont{and}
  \bibinfo{author}{\bibfnamefont{P.}~\bibnamefont{Walther}},
  \bibinfo{journal}{Nature physics} \textbf{\bibinfo{volume}{9}},
  \bibinfo{pages}{727} (\bibinfo{year}{2013}).

\bibitem[{\citenamefont{Mahadev}(2018)}]{mahadev2018classical}
\bibinfo{author}{\bibfnamefont{U.}~\bibnamefont{Mahadev}}, in
  \emph{\bibinfo{booktitle}{2018 IEEE 59th Annual Symposium on Foundations of
  Computer Science (FOCS)}} (\bibinfo{organization}{IEEE},
  \bibinfo{year}{2018}), pp. \bibinfo{pages}{259--267}.

\bibitem[{IBM({\natexlab{a}})}]{IBM}
\emph{\bibinfo{title}{\href{http://www.research.ibm.com/quantum}{{IBM} Quantum
  Experience}}}.

\bibitem[{Rig()}]{Rigetti}
\emph{\bibinfo{title}{\href{https://www.rigetti.com/}{Rigetti Computing}}}.

\bibitem[{\citenamefont{Greganti et~al.}(2018)\citenamefont{Greganti,
  Schiansky, Calafell, Procopio, Rozema, and Walther}}]{TunSPS2018}
\bibinfo{author}{\bibfnamefont{C.}~\bibnamefont{Greganti}},
  \bibinfo{author}{\bibfnamefont{P.}~\bibnamefont{Schiansky}},
  \bibinfo{author}{\bibfnamefont{I.~A.} \bibnamefont{Calafell}},
  \bibinfo{author}{\bibfnamefont{L.~M.} \bibnamefont{Procopio}},
  \bibinfo{author}{\bibfnamefont{L.~A.} \bibnamefont{Rozema}},
  \bibnamefont{and} \bibinfo{author}{\bibfnamefont{P.}~\bibnamefont{Walther}},
  \bibinfo{journal}{Opt. Expr.} \textbf{\bibinfo{volume}{26}},
  \bibinfo{pages}{3286} (\bibinfo{year}{2018}).

\bibitem[{\citenamefont{Elben et~al.}(2019)\citenamefont{Elben, Vermersch, van
  Bijnen, Kokail, Brydges, Maier, Joshi, Blatt, Roos, and Zoller}}]{Elben2019}
\bibinfo{author}{\bibfnamefont{A.}~\bibnamefont{Elben}},
  \bibinfo{author}{\bibfnamefont{B.}~\bibnamefont{Vermersch}},
  \bibinfo{author}{\bibfnamefont{R.}~\bibnamefont{van Bijnen}},
  \bibinfo{author}{\bibfnamefont{C.}~\bibnamefont{Kokail}},
  \bibinfo{author}{\bibfnamefont{T.}~\bibnamefont{Brydges}},
  \bibinfo{author}{\bibfnamefont{C.}~\bibnamefont{Maier}},
  \bibinfo{author}{\bibfnamefont{M.}~\bibnamefont{Joshi}},
  \bibinfo{author}{\bibfnamefont{R.}~\bibnamefont{Blatt}},
  \bibinfo{author}{\bibfnamefont{C.~F.} \bibnamefont{Roos}}, \bibnamefont{and}
  \bibinfo{author}{\bibfnamefont{P.}~\bibnamefont{Zoller}},
  \bibinfo{journal}{arxiv:1909.01282}  (\bibinfo{year}{2019}).

\bibitem[{\citenamefont{Raussendorf and Briegel}(2001)}]{Raussendorf2001}
\bibinfo{author}{\bibfnamefont{R.}~\bibnamefont{Raussendorf}} \bibnamefont{and}
  \bibinfo{author}{\bibfnamefont{H.~J.} \bibnamefont{Briegel}},
  \bibinfo{journal}{Phys. Rev. Lett.} \textbf{\bibinfo{volume}{86}},
  \bibinfo{pages}{5188} (\bibinfo{year}{2001}).

\bibitem[{\citenamefont{Briegel et~al.}(2009)\citenamefont{Briegel, Browne, W.,
  Raussendorf, and Van~den Nest}}]{BriegelMBQC2009}
\bibinfo{author}{\bibfnamefont{H.-J.} \bibnamefont{Briegel}},
  \bibinfo{author}{\bibfnamefont{D.~E.} \bibnamefont{Browne}},
  \bibinfo{author}{\bibfnamefont{D.}~\bibnamefont{W.}},
  \bibinfo{author}{\bibfnamefont{R.}~\bibnamefont{Raussendorf}},
  \bibnamefont{and} \bibinfo{author}{\bibfnamefont{M.}~\bibnamefont{Van~den
  Nest}}, \bibinfo{journal}{Nat. Phys.} \textbf{\bibinfo{volume}{19}},
  \bibinfo{pages}{5} (\bibinfo{year}{2009}).

\bibitem[{\citenamefont{Fitzsimons}(2017)}]{fitzsimons2017private}
\bibinfo{author}{\bibfnamefont{J.~F.} \bibnamefont{Fitzsimons}},
  \bibinfo{journal}{{npj} Quantum Information} \textbf{\bibinfo{volume}{3}},
  \bibinfo{pages}{23} (\bibinfo{year}{2017}).

\bibitem[{\citenamefont{Mantri et~al.}(2017{\natexlab{a}})\citenamefont{Mantri,
  Demarie, and Fitzsimons}}]{mantri2016universality}
\bibinfo{author}{\bibfnamefont{A.}~\bibnamefont{Mantri}},
  \bibinfo{author}{\bibfnamefont{T.~F.} \bibnamefont{Demarie}},
  \bibnamefont{and} \bibinfo{author}{\bibfnamefont{J.~F.}
  \bibnamefont{Fitzsimons}}, \bibinfo{journal}{Sci. Rep.}
  \textbf{\bibinfo{volume}{7}} (\bibinfo{year}{2017}{\natexlab{a}}).

\bibitem[{\citenamefont{Danos and Kashefi}(2006)}]{Danos2006}
\bibinfo{author}{\bibfnamefont{V.}~\bibnamefont{Danos}} \bibnamefont{and}
  \bibinfo{author}{\bibfnamefont{E.}~\bibnamefont{Kashefi}},
  \bibinfo{journal}{Phys. Rev. A} \textbf{\bibinfo{volume}{74}},
  \bibinfo{pages}{052310} (\bibinfo{year}{2006}).

\bibitem[{\citenamefont{Browne et~al.}(2007)\citenamefont{Browne, Kashefi,
  Mhalla, and Perdrix}}]{Browne2007}
\bibinfo{author}{\bibfnamefont{D.}~\bibnamefont{Browne}},
  \bibinfo{author}{\bibfnamefont{E.}~\bibnamefont{Kashefi}},
  \bibinfo{author}{\bibfnamefont{M.}~\bibnamefont{Mhalla}}, \bibnamefont{and}
  \bibinfo{author}{\bibfnamefont{S.}~\bibnamefont{Perdrix}},
  \bibinfo{journal}{New J. Phys.} \textbf{\bibinfo{volume}{9}},
  \bibinfo{pages}{250} (\bibinfo{year}{2007}).

\bibitem[{\citenamefont{Mantri et~al.}(2017{\natexlab{b}})\citenamefont{Mantri,
  Demarie, Menicucci, and Fitzsimons}}]{Mantri2016prx}
\bibinfo{author}{\bibfnamefont{A.}~\bibnamefont{Mantri}},
  \bibinfo{author}{\bibfnamefont{T.~F.} \bibnamefont{Demarie}},
  \bibinfo{author}{\bibfnamefont{N.~C.} \bibnamefont{Menicucci}},
  \bibnamefont{and} \bibinfo{author}{\bibfnamefont{J.~F.}
  \bibnamefont{Fitzsimons}}, \bibinfo{journal}{Phys. Rev. X}
  \textbf{\bibinfo{volume}{7}}, \bibinfo{pages}{031004}
  (\bibinfo{year}{2017}{\natexlab{b}}).

\bibitem[{\citenamefont{Mhalla et~al.}(2011)\citenamefont{Mhalla, Murao,
  Perdrix, Someya, and Turner}}]{mhalla2011graph}
\bibinfo{author}{\bibfnamefont{M.}~\bibnamefont{Mhalla}},
  \bibinfo{author}{\bibfnamefont{M.}~\bibnamefont{Murao}},
  \bibinfo{author}{\bibfnamefont{S.}~\bibnamefont{Perdrix}},
  \bibinfo{author}{\bibfnamefont{M.}~\bibnamefont{Someya}}, \bibnamefont{and}
  \bibinfo{author}{\bibfnamefont{P.~S.} \bibnamefont{Turner}}, in
  \emph{\bibinfo{booktitle}{Conference on Quantum Computation, Communication,
  and Cryptography}} (\bibinfo{organization}{Springer}, \bibinfo{year}{2011}),
  pp. \bibinfo{pages}{174--187}.

\bibitem[{\citenamefont{Knight and Bloom}(1973)}]{birthday}
\bibinfo{author}{\bibfnamefont{W.}~\bibnamefont{Knight}} \bibnamefont{and}
  \bibinfo{author}{\bibfnamefont{D.~M.} \bibnamefont{Bloom}},
  \bibinfo{journal}{Am. Math. Mon.} \textbf{\bibinfo{volume}{80}},
  \bibinfo{pages}{1141} (\bibinfo{year}{1973}).

\bibitem[{\citenamefont{Boixo et~al.}(2018)\citenamefont{Boixo, Isakov,
  Smelyanskiy, Babbush, Ding, Jiang, Bremner, Martinis, and Neven}}]{Boixo2018}
\bibinfo{author}{\bibfnamefont{S.}~\bibnamefont{Boixo}},
  \bibinfo{author}{\bibfnamefont{S.~V.} \bibnamefont{Isakov}},
  \bibinfo{author}{\bibfnamefont{V.~N.} \bibnamefont{Smelyanskiy}},
  \bibinfo{author}{\bibfnamefont{R.}~\bibnamefont{Babbush}},
  \bibinfo{author}{\bibfnamefont{N.}~\bibnamefont{Ding}},
  \bibinfo{author}{\bibfnamefont{Z.}~\bibnamefont{Jiang}},
  \bibinfo{author}{\bibfnamefont{M.~J.} \bibnamefont{Bremner}},
  \bibinfo{author}{\bibfnamefont{J.~M.} \bibnamefont{Martinis}},
  \bibnamefont{and} \bibinfo{author}{\bibfnamefont{H.}~\bibnamefont{Neven}},
  \bibinfo{journal}{Nat. Phys.} \textbf{\bibinfo{volume}{14}},
  \bibinfo{pages}{595} (\bibinfo{year}{2018}).

\bibitem[{\citenamefont{Pednault et~al.}(2019)\citenamefont{Pednault, Gunnels,
  Nannicini, Horesh, and Wisnieff}}]{Pednault2019}
\bibinfo{author}{\bibfnamefont{E.}~\bibnamefont{Pednault}},
  \bibinfo{author}{\bibfnamefont{J.~A.} \bibnamefont{Gunnels}},
  \bibinfo{author}{\bibfnamefont{G.}~\bibnamefont{Nannicini}},
  \bibinfo{author}{\bibfnamefont{L.}~\bibnamefont{Horesh}}, \bibnamefont{and}
  \bibinfo{author}{\bibfnamefont{R.}~\bibnamefont{Wisnieff}},
  \bibinfo{journal}{arXiv:1910.09534}  (\bibinfo{year}{2019}).

\bibitem[{\citenamefont{Linke et~al.}(2017)\citenamefont{Linke, Maslov,
  Roetteler, Debnath, Figgatt, Landsman, Wright, and Monroe}}]{Linke2017}
\bibinfo{author}{\bibfnamefont{N.~M.} \bibnamefont{Linke}},
  \bibinfo{author}{\bibfnamefont{D.}~\bibnamefont{Maslov}},
  \bibinfo{author}{\bibfnamefont{M.}~\bibnamefont{Roetteler}},
  \bibinfo{author}{\bibfnamefont{S.}~\bibnamefont{Debnath}},
  \bibinfo{author}{\bibfnamefont{C.}~\bibnamefont{Figgatt}},
  \bibinfo{author}{\bibfnamefont{K.~A.} \bibnamefont{Landsman}},
  \bibinfo{author}{\bibfnamefont{K.}~\bibnamefont{Wright}}, \bibnamefont{and}
  \bibinfo{author}{\bibfnamefont{C.}~\bibnamefont{Monroe}},
  \bibinfo{journal}{Proc Natl. Acad. Sci.} \textbf{\bibinfo{volume}{114}},
  \bibinfo{pages}{3305} (\bibinfo{year}{2017}).

\bibitem[{\citenamefont{Barz et~al.}(2012)\citenamefont{Barz, Kashefi,
  Broadbent, Fitzsimons, Zeilinger, and Walther}}]{Barz2012}
\bibinfo{author}{\bibfnamefont{S.}~\bibnamefont{Barz}},
  \bibinfo{author}{\bibfnamefont{E.}~\bibnamefont{Kashefi}},
  \bibinfo{author}{\bibfnamefont{A.}~\bibnamefont{Broadbent}},
  \bibinfo{author}{\bibfnamefont{J.}~\bibnamefont{Fitzsimons}},
  \bibinfo{author}{\bibfnamefont{A.}~\bibnamefont{Zeilinger}},
  \bibnamefont{and} \bibinfo{author}{\bibfnamefont{P.}~\bibnamefont{Walther}},
  \bibinfo{journal}{Science} \textbf{\bibinfo{volume}{335}},
  \bibinfo{pages}{303} (\bibinfo{year}{2012}).

\bibitem[{\citenamefont{Lanyon et~al.}(2013)\citenamefont{Lanyon, Jurcevic,
  Zwerger, Hempel, Martinez, D\"ur., Briegel, Blatt, and Roos}}]{Lanyon2013}
\bibinfo{author}{\bibfnamefont{B.~P.} \bibnamefont{Lanyon}},
  \bibinfo{author}{\bibfnamefont{P.}~\bibnamefont{Jurcevic}},
  \bibinfo{author}{\bibfnamefont{M.}~\bibnamefont{Zwerger}},
  \bibinfo{author}{\bibfnamefont{C.}~\bibnamefont{Hempel}},
  \bibinfo{author}{\bibfnamefont{E.~A.} \bibnamefont{Martinez}},
  \bibinfo{author}{\bibnamefont{D\"ur.}}, \bibinfo{author}{\bibfnamefont{H.~J.}
  \bibnamefont{Briegel}},
  \bibinfo{author}{\bibfnamefont{R.}~\bibnamefont{Blatt}}, \bibnamefont{and}
  \bibinfo{author}{\bibfnamefont{C.~F.} \bibnamefont{Roos}},
  \bibinfo{journal}{Phys. Rev. Lett.} p. \bibinfo{pages}{210501}
  (\bibinfo{year}{2013}).

\bibitem[{\citenamefont{McCutcheon et~al.}(2018)\citenamefont{McCutcheon,
  McMillan, Rarity, and Tame}}]{Cutcheon2018}
\bibinfo{author}{\bibfnamefont{W.}~\bibnamefont{McCutcheon}},
  \bibinfo{author}{\bibfnamefont{A.}~\bibnamefont{McMillan}},
  \bibinfo{author}{\bibfnamefont{J.~G.} \bibnamefont{Rarity}},
  \bibnamefont{and} \bibinfo{author}{\bibfnamefont{M.~S.} \bibnamefont{Tame}},
  \bibinfo{journal}{New Journal of Physics} \textbf{\bibinfo{volume}{20}}
  (\bibinfo{year}{2018}).

\bibitem[{\citenamefont{Kawamura et~al.}(2010)\citenamefont{Kawamura,
  B.~Rowland, and Jones}}]{Kawamura2010}
\bibinfo{author}{\bibfnamefont{M.}~\bibnamefont{Kawamura}},
  \bibinfo{author}{\bibfnamefont{B.}~\bibnamefont{B.~Rowland}},
  \bibnamefont{and} \bibinfo{author}{\bibfnamefont{J.~A.} \bibnamefont{Jones}},
  \bibinfo{journal}{Phys. Rev. A} \textbf{\bibinfo{volume}{82}},
  \bibinfo{pages}{032315} (\bibinfo{year}{2010}).

\bibitem[{\citenamefont{Knill et~al.}(2000)\citenamefont{Knill, Laflamme,
  Martinez, and H.}}]{Knill2000}
\bibinfo{author}{\bibfnamefont{E.}~\bibnamefont{Knill}},
  \bibinfo{author}{\bibfnamefont{R.}~\bibnamefont{Laflamme}},
  \bibinfo{author}{\bibfnamefont{R.}~\bibnamefont{Martinez}}, \bibnamefont{and}
  \bibinfo{author}{\bibfnamefont{T.~C.} \bibnamefont{H.}},
  \bibinfo{journal}{Nature} \textbf{\bibinfo{volume}{404}},
  \bibinfo{pages}{368} (\bibinfo{year}{2000}).

\bibitem[{\citenamefont{Bowdrey et~al.}(2005)\citenamefont{Bowdrey, Jones,
  Knill, and Laflamme}}]{Bowdrey2005}
\bibinfo{author}{\bibfnamefont{M.~D.} \bibnamefont{Bowdrey}},
  \bibinfo{author}{\bibfnamefont{J.~A.} \bibnamefont{Jones}},
  \bibinfo{author}{\bibfnamefont{E.}~\bibnamefont{Knill}}, \bibnamefont{and}
  \bibinfo{author}{\bibfnamefont{R.}~\bibnamefont{Laflamme}},
  \bibinfo{journal}{Phys. Rev. A} \textbf{\bibinfo{volume}{72}},
  \bibinfo{pages}{032315} (\bibinfo{year}{2005}).

\bibitem[{\citenamefont{Lu et~al.}(2007)\citenamefont{Lu, Zhou, O., Gao, Zhang,
  .~Yuan, Goebel, Yang, and Pan}}]{Lu2006}
\bibinfo{author}{\bibfnamefont{C.-Y.} \bibnamefont{Lu}},
  \bibinfo{author}{\bibfnamefont{X.-Q.} \bibnamefont{Zhou}},
  \bibinfo{author}{\bibfnamefont{G.}~\bibnamefont{O.}},
  \bibinfo{author}{\bibfnamefont{W.-B.} \bibnamefont{Gao}},
  \bibinfo{author}{\bibfnamefont{J.}~\bibnamefont{Zhang}},
  \bibinfo{author}{\bibfnamefont{Z.-S.} \bibnamefont{.~Yuan}},
  \bibinfo{author}{\bibfnamefont{A.}~\bibnamefont{Goebel}},
  \bibinfo{author}{\bibfnamefont{T.}~\bibnamefont{Yang}}, \bibnamefont{and}
  \bibinfo{author}{\bibfnamefont{J.-W.} \bibnamefont{Pan}},
  \bibinfo{journal}{Nat. Phys.} \textbf{\bibinfo{volume}{3}},
  \bibinfo{pages}{91} (\bibinfo{year}{2007}).

\bibitem[{\citenamefont{Saggio et~al.}(2019)\citenamefont{Saggio, Dimi\'c,
  Greganti, Rozema, Walther, and Daki\'c}}]{SingleCopy}
\bibinfo{author}{\bibfnamefont{V.}~\bibnamefont{Saggio}},
  \bibinfo{author}{\bibfnamefont{A.}~\bibnamefont{Dimi\'c}},
  \bibinfo{author}{\bibfnamefont{C.}~\bibnamefont{Greganti}},
  \bibinfo{author}{\bibfnamefont{L.~A.} \bibnamefont{Rozema}},
  \bibinfo{author}{\bibfnamefont{P.}~\bibnamefont{Walther}}, \bibnamefont{and}
  \bibinfo{author}{\bibfnamefont{B.}~\bibnamefont{Daki\'c}},
  \bibinfo{journal}{Nat. Phys.} \textbf{\bibinfo{volume}{15}},
  \bibinfo{pages}{935–940} (\bibinfo{year}{2019}).

\bibitem[{\citenamefont{Broome et~al.}(2011)\citenamefont{Broome, Almeida,
  Fedrizzi, and White}}]{broome11}
\bibinfo{author}{\bibfnamefont{M.~A.} \bibnamefont{Broome}},
  \bibinfo{author}{\bibfnamefont{M.~P.} \bibnamefont{Almeida}},
  \bibinfo{author}{\bibfnamefont{A.}~\bibnamefont{Fedrizzi}}, \bibnamefont{and}
  \bibinfo{author}{\bibfnamefont{A.~G.} \bibnamefont{White}},
  \bibinfo{journal}{Opt. Expr.} \textbf{\bibinfo{volume}{19}},
  \bibinfo{pages}{22698} (\bibinfo{year}{2011}).

\bibitem[{\citenamefont{Greganti et~al.}(2015)\citenamefont{Greganti, Roehsner,
  Barz, Waegel, and Walther}}]{Mordecai}
\bibinfo{author}{\bibfnamefont{C.}~\bibnamefont{Greganti}},
  \bibinfo{author}{\bibfnamefont{M.~C.} \bibnamefont{Roehsner}},
  \bibinfo{author}{\bibfnamefont{S.}~\bibnamefont{Barz}},
  \bibinfo{author}{\bibfnamefont{M.}~\bibnamefont{Waegel}}, \bibnamefont{and}
  \bibinfo{author}{\bibfnamefont{P.}~\bibnamefont{Walther}},
  \bibinfo{journal}{Phys. Rev. A} \textbf{\bibinfo{volume}{91}},
  \bibinfo{pages}{022325} (\bibinfo{year}{2015}).

\bibitem[{\citenamefont{Dimi{\'{c}} and Daki{\'{c}}}(2018)}]{SingleCopy2}
\bibinfo{author}{\bibfnamefont{A.}~\bibnamefont{Dimi{\'{c}}}} \bibnamefont{and}
  \bibinfo{author}{\bibfnamefont{B.}~\bibnamefont{Daki{\'{c}}}},
  \bibinfo{journal}{npj Quantum Information} \textbf{\bibinfo{volume}{4}},
  \bibinfo{pages}{11} (\bibinfo{year}{2018}).

\bibitem[{IBM({\natexlab{b}})}]{IBMqx4}
\emph{\bibinfo{title}{5-qubit backend: {IBM} {Q} team, ``{IBM} {Q} 5 yorktown
  backend specification {V}1.0.0,'' (2017). retrieved from
  \href{https://ibm.biz/qiskit-yorktown}{https://ibm.biz/qiskit-yorktown}.}}

\bibitem[{\citenamefont{Otterbach et~al.}(2017)\citenamefont{Otterbach,
  Manenti, Alidoust, Bestwick, Block, Bloom, Caldwell, Didier, Schuyler~Fried,
  Hong et~al.}}]{Rigetti19Q}
\bibinfo{author}{\bibfnamefont{J.~S.} \bibnamefont{Otterbach}},
  \bibinfo{author}{\bibfnamefont{R.}~\bibnamefont{Manenti}},
  \bibinfo{author}{\bibfnamefont{N.}~\bibnamefont{Alidoust}},
  \bibinfo{author}{\bibfnamefont{A.}~\bibnamefont{Bestwick}},
  \bibinfo{author}{\bibfnamefont{M.}~\bibnamefont{Block}},
  \bibinfo{author}{\bibfnamefont{B.}~\bibnamefont{Bloom}},
  \bibinfo{author}{\bibfnamefont{S.}~\bibnamefont{Caldwell}},
  \bibinfo{author}{\bibfnamefont{N.}~\bibnamefont{Didier}},
  \bibinfo{author}{\bibfnamefont{E.}~\bibnamefont{Schuyler~Fried}},
  \bibinfo{author}{\bibfnamefont{S.}~\bibnamefont{Hong}}, \bibnamefont{et~al.},
  \bibinfo{journal}{arXiv:1712.05771}  (\bibinfo{year}{2017}).

\bibitem[{\citenamefont{Wendin}(2017)}]{Wendin2017}
\bibinfo{author}{\bibfnamefont{G.}~\bibnamefont{Wendin}},
  \bibinfo{journal}{Rep. Prog. Phys.} \textbf{\bibinfo{volume}{80}},
  \bibinfo{pages}{10} (\bibinfo{year}{2017}).

\bibitem[{\citenamefont{Caldwell et~al.}(2018)\citenamefont{Caldwell, Didier,
  Ryan, Sete, Hudson, Karalekas, Manenti, da~Silva, Sinclair, Acala
  et~al.}}]{Rigetti2017}
\bibinfo{author}{\bibfnamefont{S.~A.} \bibnamefont{Caldwell}},
  \bibinfo{author}{\bibfnamefont{N.}~\bibnamefont{Didier}},
  \bibinfo{author}{\bibfnamefont{C.~A.} \bibnamefont{Ryan}},
  \bibinfo{author}{\bibfnamefont{E.~A.} \bibnamefont{Sete}},
  \bibinfo{author}{\bibfnamefont{A.}~\bibnamefont{Hudson}},
  \bibinfo{author}{\bibfnamefont{P.}~\bibnamefont{Karalekas}},
  \bibinfo{author}{\bibfnamefont{R.}~\bibnamefont{Manenti}},
  \bibinfo{author}{\bibfnamefont{M.~P.} \bibnamefont{da~Silva}},
  \bibinfo{author}{\bibfnamefont{R.}~\bibnamefont{Sinclair}},
  \bibinfo{author}{\bibfnamefont{E.}~\bibnamefont{Acala}},
  \bibnamefont{et~al.}, \bibinfo{journal}{Phys. Rev. Appl.}
  \textbf{\bibinfo{volume}{10}}, \bibinfo{pages}{034050}
  (\bibinfo{year}{2018}).

\bibitem[{\citenamefont{Didier et~al.}(2018)\citenamefont{Didier, Sete,
  da~Silva, and Rigetti}}]{Rigetti2018}
\bibinfo{author}{\bibfnamefont{N.}~\bibnamefont{Didier}},
  \bibinfo{author}{\bibfnamefont{E.~A.} \bibnamefont{Sete}},
  \bibinfo{author}{\bibfnamefont{M.~P.} \bibnamefont{da~Silva}},
  \bibnamefont{and} \bibinfo{author}{\bibfnamefont{C.}~\bibnamefont{Rigetti}},
  \bibinfo{journal}{Phys. Rev. A} \textbf{\bibinfo{volume}{97}},
  \bibinfo{pages}{022330} (\bibinfo{year}{2018}).

\bibitem[{\citenamefont{Sorensen and Molmer}(2000)}]{Sorensen2000}
\bibinfo{author}{\bibfnamefont{A.}~\bibnamefont{Sorensen}} \bibnamefont{and}
  \bibinfo{author}{\bibfnamefont{K.}~\bibnamefont{Molmer}},
  \bibinfo{journal}{Phys. Rev. A} \textbf{\bibinfo{volume}{62}},
  \bibinfo{pages}{022311} (\bibinfo{year}{2000}).

\bibitem[{\citenamefont{Wang et~al.}(2016)\citenamefont{Wang, Chen, Li, Huang,
  Liu, Chen, Luo, Su, Wu, Li et~al.}}]{pan10}
\bibinfo{author}{\bibfnamefont{X.-L.} \bibnamefont{Wang}},
  \bibinfo{author}{\bibfnamefont{L.-K.} \bibnamefont{Chen}},
  \bibinfo{author}{\bibfnamefont{W.}~\bibnamefont{Li}},
  \bibinfo{author}{\bibfnamefont{H.-L.} \bibnamefont{Huang}},
  \bibinfo{author}{\bibfnamefont{C.}~\bibnamefont{Liu}},
  \bibinfo{author}{\bibfnamefont{C.}~\bibnamefont{Chen}},
  \bibinfo{author}{\bibfnamefont{Y.-H.} \bibnamefont{Luo}},
  \bibinfo{author}{\bibfnamefont{Z.-E.} \bibnamefont{Su}},
  \bibinfo{author}{\bibfnamefont{D.}~\bibnamefont{Wu}},
  \bibinfo{author}{\bibfnamefont{Z.-D.} \bibnamefont{Li}},
  \bibnamefont{et~al.}, \bibinfo{journal}{Phys. Rev. Lett.}
  \textbf{\bibinfo{volume}{117}} (\bibinfo{year}{2016}).

\bibitem[{\citenamefont{Gottesman and Chuang}(1999)}]{Gottesman1999}
\bibinfo{author}{\bibfnamefont{D.}~\bibnamefont{Gottesman}} \bibnamefont{and}
  \bibinfo{author}{\bibfnamefont{I.~L.} \bibnamefont{Chuang}},
  \bibinfo{journal}{Nature} \textbf{\bibinfo{volume}{402}},
  \bibinfo{pages}{390} (\bibinfo{year}{1999}).

\bibitem[{\citenamefont{Gottesman}(1997)}]{Gottesman1997}
\bibinfo{author}{\bibfnamefont{D.}~\bibnamefont{Gottesman}}, Ph.D. thesis,
  \bibinfo{school}{California Institute of Technology},
  \bibinfo{address}{Pasadena, CA} (\bibinfo{year}{1997}).

\bibitem[{ibm()}]{ibmqx3}
\emph{\bibinfo{title}{16-qubit backend: {IBM} {Q} team, "{IBM} {Q} 16
  rueschlikon backend specification {V}1.0.0," (2017). retrieved from
  \href{https://ibm.biz/qiskit-rueschlikon}{https://ibm.biz/qiskit-rueschlikon}.}}

\end{thebibliography}

\section*{Acknowledgements}
The authors thank Sam Ferracin for helpful comments on the manuscript.\\
\textbf{Funding} We gratefully acknowledge support by the Austrian Science Fund (FWF), through the SFB FoQuS (FWF Project No.\ F40) and SFB BeyondC (FWF Project No.\ F71). We also acknowledge funding from the EU H2020-FETFLAG-2018-03 under Grant Agreement no. 820495, by the U.S. Army Research Office (ARO) through grant no. W911NF-14-1-0103, and by the Office of the Director of National Intelligence (ODNI), Intelligence Advanced Research Projects Activity (IARPA), via the U.S.\ ARO Grant No.\ W911NF-16-1-0070. This project has received funding from the European Union’s Horizon 2020 research and innovation programme under the Marie Skłodowska-Curie grant agreement No 801110 and the Austrian Federal Ministry of Education, Science and Research (BMBWF). It reflects only the author's view, the EU Agency is not responsible for any use that may be made of the information it contains. PW acknowledges support from the research platform TURIS, the Austrian Science Fund (FWF) through the Doctoral Programme CoQuS (no. W1210-4), BeyondC (F71) and NaMuG (P30067-N36), the U.S.\ Air Force Office of Scientific Research via QAT4SECOMP (FA2386-17-1-4011), the European Commission via High dimensional quantum Photonic Platform projects (HiPhoP) (no.\ 731473) and Red Bull GmbH. JFF acknowledges support from Singapore's Ministry of Education, National Research Foundation (ANR-NRF Grant No.\ NRF2017-NRF-ANR004), and the U.S.\ Air Force Office of Scientific Research (AOARD Grant No.\ FA2386-18-1-4003). \\
\textbf{Author Contributions} JFF and TFD conceived the project and derived the theory results. CG, VS, IAC, and LAR performed the photonic experiments; MR, AE, MM, LP, RS, PS, and TM performed the trapped ion experiments; JAJ performed the NMR experiments; TFD programmed the superconducting experiments. CG, TFD and MR analyzed the data. RB, TM, PW, and JFF supervised the project. All authors contributed to writing the manuscript. \\
\textbf{Competing interests} The authors declare no competing interests. None of the affiliated commercially-oriented companies have been partners or collaborators in the context of this scientific work.
\textbf{Data and materials availability:} Correspondence and requests for materials should be addressed to JFF (email: joe@horizonquantum.com).

\normalsize

\newpage
\onecolumngrid
\clearpage
\renewcommand{\theequation}{S\arabic{equation}}
\renewcommand{\thefigure}{S\arabic{figure}}
\renewcommand{\thetable}{S\Roman{table}}
\renewcommand{\thesection}{S\Roman{section}}
\setcounter{equation}{0}
\setcounter{figure}{0}
\setcounter{table}{0}
\setcounter{section}{0}
\begin{center}
{\bf \large Supplementary Information: \\
Cross-verification of independent quantum devices}
\end{center}
\medskip
Here we provide full details on the conversion between MBQC and circuit model computations for different choices of flow. We illustrate this using an explicit example from the main text and also provide complementary results on other graph states. Finally, we discuss some additional experimental details of the single-photon implementation.
\medskip
\twocolumngrid

\section{Experimental Methods}
\textbf{NMR \emph{(Oxford)}} experiments~\cite{Jones2011} were performed on a Varian Unity Inova spectrometer with a nominal \nuc{1}{H} frequency of \SI{600}{\MHz} using a H\{CN\} probe with a single pulsed field gradient. The NMR sample comprised \nuc{13}{C}-labelled sodium formate dissolved in $\textrm{D}_2\textrm{O}$ at \SI{25}{\degree}C, providing a heteronuclear two-spin system. With both spins on resonance, the Hamiltonian took the form of a spin--spin $ZZ$ coupling of \SI{194.7}{\Hz}, and the \Bone\ field strengths were measured to give nutation rates of approximately \SI{25}{\kHz} for \nuc{1}{H} and \SI{17}{\kHz} for \nuc{13}{C}.

Pseudo-pure two-qubit states were prepared using the method of Ref.~\onlinecite{Kawamura2010}. Single-qubit rotations in the $XY$-plane were implemented using simple pulses, while two-qubit rotations were implemented as delays. Fixed $Z$-rotations were implemented as frame rotations~\cite{Knill2000} which were propagated through the pulse sequence~\cite{Bowdrey2005} to points where they could be dropped. The variable small-angle $Z$-rotations were implemented using a pair of $\pi$ pulses with phases separated by $\theta/2$, with the phase of the first pulse chosen to partially cancel with the preceding Hadamard gate. 

At the end of the algorithm a crush gradient was applied to project the density matrix onto the computational basis, and the \nuc{1}{H} NMR spectrum observed after a $\pi/2$ pulse. NMR spectra were processed using custom software and the intensity of the two components of the \nuc{1}{H} doublet were determined by integration and normalized to a reference spectrum. Corresponding measurements on the second qubit were performed by repeating the experiment with the reverse assignment of qubits to physical spins. From the collection of these measurements the populations of the computational basis states can be estimated. Due to imperfect calibration these populations do not quite sum to one, and some can be slightly negative. This was resolved by subtracting the most negative population found in any group of experiments from all the populations in that group, and then normalizing the populations for each experiment.

\textbf{Photons \emph{(Vienna)}} experiments are based on the generation of the maximally-entangled six-qubit H-shaped cluster state. Three polarization-entangled pairs of photons are produced via three identical Sagnac-PPKTP pulsed down-conversion sources and later entangled by using partial fusion gates at polarizing beam splitters~\cite{Lu2006, SingleCopy}. The qubits are encoded by the polarization of the six photons. 

The laser repetition rate is set to \SI{152}{\MHz}, by doubling the original rate with a passive multiplexing scheme~\cite{broome11}, to reduce multi-photon noise for an average power of $220$ mW per source. The pump photons have a wavelength of \SI{772.9}{\nm} and a pulse-width of \SI{2.1}{\ps}. The crystals' temperature is stabilized at \SI{24}{\celsius}. The single-qubit measurements are implemented with an optics tomographic unit of three motorized waveplates and a polarizing beam splitter per photon. Twelve multi-element superconducting-nanowire single-photon detectors, composed of 4 channels each and kept at $T=0.9$ K, enable a pseudo-number resolving detection, with an average quantum efficiency of $0.87$. Due to technical problems, two of the multi-element detectors had to be later replaced with two single-element detectors. A customized time tagging and logic module for 48 input-channels counts the six-fold photons events. After postselection we obtain a total six-fold coincidence rate of \SI{0.08}{\Hz}. The purity of the single photons, measured with four-fold HOM interference, corresponds to $0.94$~\cite{TunSPS2018}. We characterize the six-photon cluster state  by using a subset of stabilizer operators, so-called identity product \cite{Mordecai}, giving a lower bound on the state fidelity of $F_{exp}\geq 0.64\pm 0.04$ and by using a technique based on a probabilistic protocol for entanglement detection ~\cite{SingleCopy,SingleCopy2}, estimating a fidelity of $0.75 \pm 0.06$ (see SM).

\textbf{Superconducting \emph{(IBM and Rigetti)}} qubits are used independently via the two cloud-accessible quantum processors: the \emph{ibmqx2} (also known as \emph{IBM Q 5 Yorktown}) from IBM \cite{IBM,IBMqx4} and the \emph{Rigetti 19Q} from Rigetti \cite{Rigetti, Rigetti19Q}. Both apparatuses use \emph{transmon} qubits, charge qubits which show insensitivity to charge noise thanks to an additional large capacitor in the circuit. Variations of the two devices can be found on the circuit wiring and reading, and the fabrication materials, e.g.\ the ibmqx2 has a star-shaped connected circuit, based on fixed-frequency transmons \cite{Wendin2017}, with three qubits available as control qubits, whereas the Rigetti19Q has tunable-frequency transmon qubits \cite{Rigetti2017,Rigetti2018}, each coupled to three fixed-frequency transmon qubits.

The ibmqx2 processor was calibrated twice during our experimental runs and kept at a temperature of $17.5$ mK. We selected qubits 2, 3, and 4 with frequencies of $[5.2,5.0,5.3]$ GHz, single-qubit gate errors of $[3.4,3.6,3.3] \cdot 10^{-3}$, and readout errors of $[3.5,1.5,1.6] \cdot 10^{-2}$, respectively. The two-qubit gate errors consist of $[6.7,3.7]\cdot 10^{-2}$ for the controlled gate between qubit 2 and 3 and between qubit 2 and 4, respectively. 
The coherence times are $[48.5,51.7,39.4]$ $\mu$s for depolarization and $[28.9,75.6,49.9]$ $\mu$s for spin dephasing, whereas the gate time is $\sim 250$ ns. 

On the Rigetti19Q we exploit the qubits labelled 2, 8 and 13, as they show reduced noise. The chip is maintained at a temperature of $10$ mK. From \cite{Rigetti2018}, single-qubit readout fidelities are equal to $0.97$, $0.947$, and $0.921$, single-qubit gate fidelities correspond to $0.981$, $0.987$ and $0.993$ for qubits 2, 8 and 13, respectively, and two-qubit gate fidelities are $0.906$ (between qubit 2 and 8) and $0.881$ (between qubit 8 and 13). The qubits coherence time is $\sim$ $20$ $\mu$s, whereas entangling gates time is \num{100}-\SI{250}{\ns}.

\textbf{Trapped ions \emph{(Innsbruck)}} experiments are performed with qubits encoded in the electronic states of a string of ${}^{40}$Ca${}^+$ ions confined in a linear Paul trap~\cite{Schindler2013}. Each ion encodes a qubit in the ground state S$_{1/2}(m=-1/2)=\ket{1}$ and the meta-stable state D$_{5/2}(m=-1/2)=\ket{0}$, which determines the qubit lifetime of $\sim$ $1$ s. Single qubit Z-rotations are implemented via Stark-shifts induced by tightly focused laser beams, while collective rotations around any equatorial axis of the Bloch-sphere are achieved by resonant illumination of the whole ion string. Entangling operations are implemented via global M{\o}lmer-S{\o}rensen interactions using a bi-chromatic laser field~\cite{Sorensen2000}. Local gates as well as two-qubit entangling gates achieve fidelities greater than $99\%$ and operate on a timescale of $20{-}30$ $\mu$s ($80$ $\mu$s for entangling gates), much faster than the coherence time which is on the order of $\sim$ $100$ ms and dominated by laser phase noise.
Every run of the experiment consists of Doppler and sideband cooling of the ion string, followed by a gate sequence, and finally projection onto the computational subspace via fluorescence detection on the P$_{1/2}$---S$_{1/2}$ transition with a CCD camera. One such run takes ${\sim}$\SI{15}{\ms} and each experiment is repeated at least 100 times to gather statistics.

\section{Photonic H-shaped cluster state characterization}
\begin{figure}[t]
\begin{center}
\includegraphics[width=1.\columnwidth]{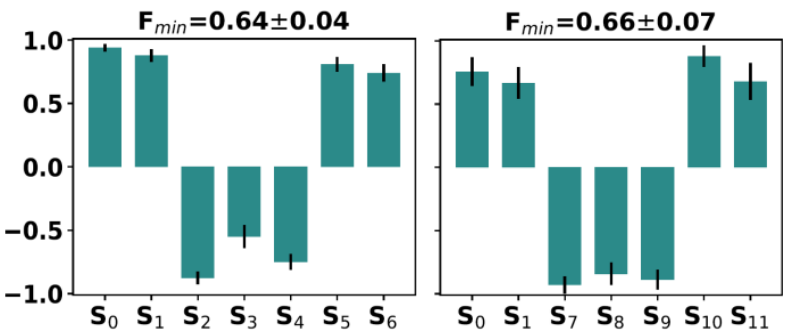} 
\caption{Expectation values of the measured stabilizer operators for the two identity product related to the 6-qubit H-shaped cluster state.}
\label{fig:FidCluster}
\end{center}
\end{figure}

We first characterize the six-photon cluster state using a technique based on subsets of stabilizer operators, referred to as Identity Products (ID)~\cite{Mordecai}. The method exploits the entanglement of the operators  to obtain a lower bound on the fidelity of the state and a proof of a Bell-type inequality  with a minimal number of measurement settings. There exist a large, unquantified number of equivalent minimal subsets of stabilizers for the 6-qubit states. Here we repeat the characterization procedure with two equivalent ID sets, composed of 7 measurements:

\begin{center}
\begin{tabular}{c}
ZZIIII  \\
ZZIXXX  \\
IZIXYY \\
XYXYYY \\
YYYYXX \\
XXYYXX \\
YXXYXX\\
(a) \\
\end{tabular}
\hspace{2.1cm}
\begin{tabular}{c}
ZZIIII \\
ZZIXXX \\
IZIYYX \\
YYYXYX\\
XXYYYY \\
XYXXYX\\
YXXXXY \\
(b) \\
\end{tabular}
\end{center}
where we have omitted the tensor product symbols for compactness. From the ID measurements we extract an ID-Bell parameter $\langle \alpha_{ID}\rangle_{exp}$, where $ \alpha_{ID}=\sum_i^M\lambda_i O_i$ and $M$ is the number of measurements settings in the ID and $O_i$ is the $i^\text{th}$ stabilizer operator of the ID. A Bell-type violation in this case is obtained if $\langle \alpha_{ID}\rangle_{exp} >M-2$. The experimental results show violations of the Bell-type inequality of $3.4\sigma$ and $2.4\sigma$, respectively. A minimum value of the fidelity can be calculated as $F_{min}=(\langle \alpha_{ID}\rangle_{exp} -M +4)/4$, providing $F_{min}=0.64 \pm 0.04$  for the first ID  and $F_{min}=0.66 \pm 0.07$ for the second ID. The error bars are reduced for the first set because of a longer acquisition time: $1.5$ h and $0.7$ h for the first and second set, respectively. The experimental expectation values for the two IDs are reported in Fig.~\ref{fig:FidCluster}. In both cases the results of the 6-qubit H-shaped cluster state show a violation of the ID-Bell inequality and high minimal fidelity. The non-ideal results are mainly due to the unbalanced losses present at the polarizing beam splitters stage, the imperfect polarization compensation along the single mode fibers connecting the three sources, and the non-unity purity of the single photons.

\begin{figure*}[t]
\begin{center}
\includegraphics[width=0.95\linewidth]{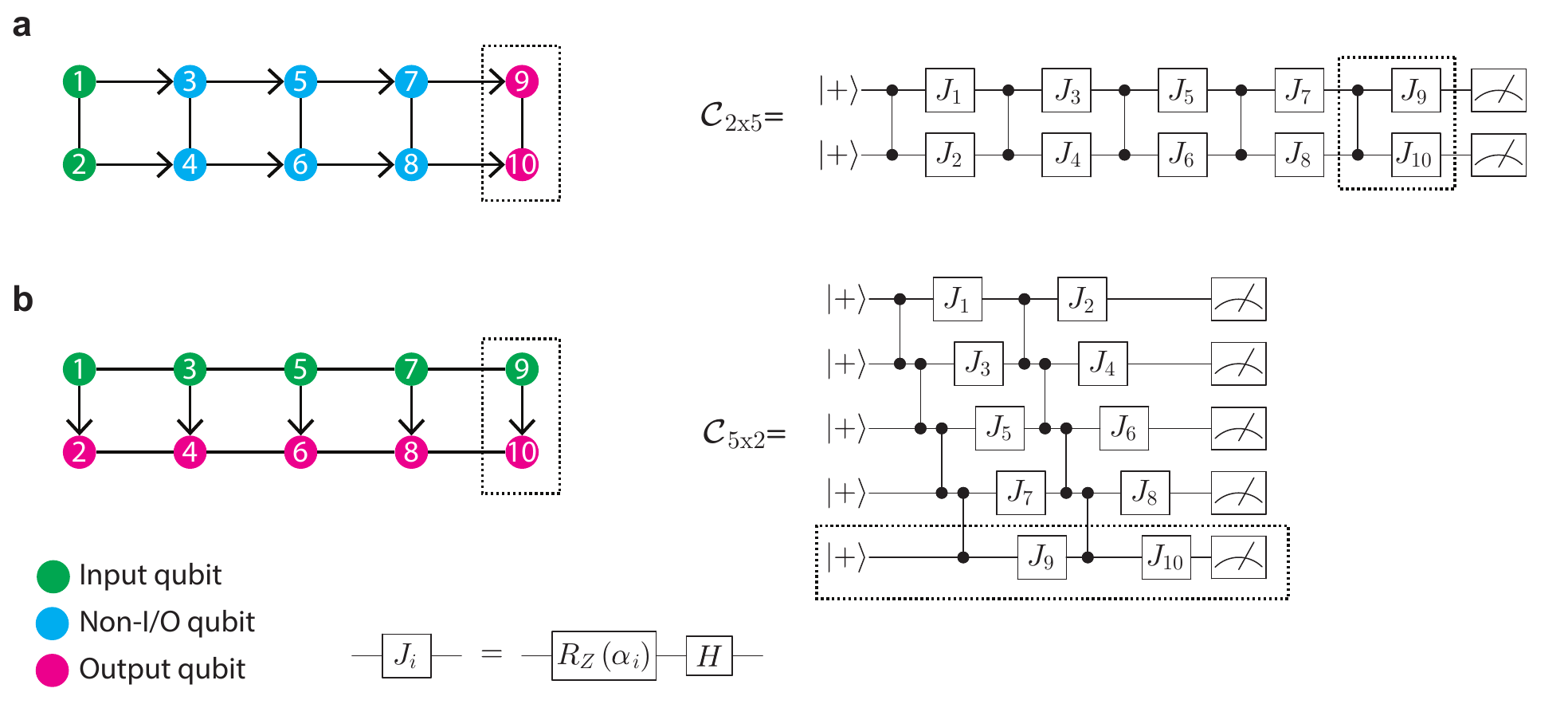} 
\caption{The Box Cluster $2$x$5$ with two choices of g-flow. \textbf{a} The ``left-to-right'' flow maps to the 2-qubit, depth-5 circuit $\mathcal{C}_{2x5}$, whereas the \textbf{b} ``top-to-bottom'' flow maps to a 5-qubit, depth-2 circuit $\mathcal{C}_{5x2}$. The construction for the Box Cluster $2$x$4$ is equivalent with the elements in dashed borders removed.}
\label{Fig:SuppBox}
\end{center}
\end{figure*}

Furthermore we follow a probabilistic protocol for entanglement detection ~\cite{SingleCopy,SingleCopy2} in order to estimate the experimental fidelity of the state. This method entails a significant reduction of resources, that is, it needs in our case only a very low number of detection events (around $100$) to verify the presence of entanglement in our cluster state with more than $99\%$ confidence.  We obtain a fidelity of $0.75 \pm 0.06$, which is comparable to fidelities obtained in state-of-the-art photonic experiments~\cite{pan10}. More details about the cluster state characterization can be found in Ref.~\cite{SingleCopy}.

\section{Converting between circuit model and MBQC}
\label{theoexample}
The reference gate in MBQC is $\hat{J}(\alpha) = \hat{H} \hat{R}_z(\alpha)$, which follows from the basics of the one-qubit teleportation scheme~\cite{Gottesman1999,mantri2016universality}. Single qubit universality is obtained by realizing that $\hat{J}(\alpha) \hat{J}(0) = \hat{R}_x(\alpha)$. 

The underlying graph for the MBQC pattern can be constructed by decomposing a generic unitary computation on a fixed initial state, $\ket{+}^{\otimes N}$, in terms of $\hat{J}(\alpha) $ gates and $\hat{C}_Z$ entangling gates. For each $\hat{J}(\alpha) $ gate we add a vertex, and draw an edge to connect this vertex to the vertex that represents the preceding $\hat{J}(\alpha) $ gate as dictated by the circuit. This is done recursively, hence creating $N$ wires, which represent the unitary evolution of each initial qubit state. The last step is drawing an edge for each $\hat{C}_Z$ gate, by connecting the two vertices representing the $\hat{J}$ gates that immediately follow the $\hat{C}_Z$ gate in the quantum circuit representation. These few steps give us the adjacency matrix of a graph $\mathcal{G} = \{ \mathcal{V}, \mathcal{E} \}$, with vertex set $\mathcal{V}$, and edge set $\mathcal{E}$. The cardinality of the vertex set is $|\mathcal{V}| = N + M$, where $M$ is the total number of $\hat{J}(\alpha)$ gates in the circuit.

\section{Circuits for the 6-qubit H-shaped cluster}
\label{example6qH}
We consider the two quantum circuits shown in Fig.~\ref{fig:2} of the main text, associated with the six-qubit H-shaped cluster state. The unitary evolution of two g-flows correspond to:
\begin{align}
\hat{\mathbf{J}}(\alpha_5,\alpha_6) \hat{\mathbf{J}}(\alpha_3,\alpha_4) C_Z  \hat{\mathbf{J}}(\alpha_1,\alpha_2) & \ket{{++}} ,\\
\hat{\mathbf{J}}(\alpha_6)_3\hat{\mathbf{J}}(\alpha_2,\alpha_4)_{1,3}C_{Z_{(1,3)}} \hat{\mathbf{J}}(\alpha_5,\alpha_3)_{2,3} C_{Z_{(2,3)}} \hat{\mathbf{J}}(\alpha_1)_3 &\ket{{+++}} ,
\end{align}
where $\hat{\mathbf{J}}(\alpha_i,\alpha_j) = J(\alpha_i)_1 \otimes J(\alpha_j)_2$ and $\hat{\mathbf{J}}(\alpha_i,\alpha_j, \alpha_k) = J(\alpha_i)_1 \otimes J(\alpha_j)_2\otimes J(\alpha_k)_3$. When using $\hat{\mathbf{J}}(\alpha_i)$ or $\hat{\mathbf{J}}(\alpha_i,\alpha_j)$ in a circuit with more qubits, this is implicitly understood as acting on the first set of qubits, unless a subscript indicates which qubits are acted on. The angles $\alpha$ can be randomly chosen within a specific set. The relationships between MBQC-related outcomes of the circuits $\mathcal{C}_{\text{a}}$ and $\mathcal{C}_{\text{b}}$ are
\begin{align}
&\text{Pr}(0,0)_{\mathcal{C}_{\text{a}}} = 2 \cdot \text{Pr}(0,0,0)_{\mathcal{C}_{\text{b}}} ,\nonumber \\ 
&\text{Pr}(0,1)_{\mathcal{C}_{\text{a}}} = 2 \cdot\text{Pr}(0,0,1)_{\mathcal{C}_{\text{b}}} ,\nonumber \\
&\text{Pr}(1,0)_{\mathcal{C}_{\text{a}}} = 2 \cdot\text{Pr}(0,1,0)_{\mathcal{C}_{\text{b}}} ,\nonumber \\
&\text{Pr}(1,1)_{\mathcal{C}_{\text{a}}} = 2 \cdot\text{Pr}(0,1,1)_{\mathcal{C}_{\text{b}}} ,
\end{align}
where the labels of the outcomes are $\text{Pr}(b_5,b_6)_{\mathcal{C}_{\text{a}}}$, and  $\text{Pr}(b_2,b_5,b_6)_{\mathcal{C}_{\text{b}}}$. 

The MBQC protocols can be expressed in terms of the stabilizer formalism \cite{Gottesman1997}. A graph state is invariant under stabilizer operations: Given a graph state on $n$ qubits $\ket{\mathcal{G}} = (\prod_{\mathcal{G}} \hat{C}_Z) \ket{+}^\otimes n$ we have: 
\begin{equation}
\hat{K}_v \ket{\mathcal{G}} = \ket{\mathcal{G}}\, , \quad \forall v \in \mathcal{V} \, .
\end{equation}
We can rewrite the computation by applying a stabilizer operator on each vertex of the graph state. We consider the stabilizers in their most general form, not restricting to the Pauli group, and a random bit-string $\textbf{k} = \{k_i\}_{i=1}^6$, $k_i \in \mathbb{Z}_2$ associated with the six stabilizers. Then the measurement angles $\alpha$ can be rewritten as
\begin{equation}
\text{angles} =
	 \begin{dcases*}
		\tilde{\alpha}_1 = (-1)^{k_1} \alpha_1 + k_3 \pi \\
		\tilde{\alpha}_2 = (-1)^{k_2} \alpha_2 + (k_4 + r_1) \pi \\
		\tilde{\alpha}_3 = (-1)^{k_3} \alpha_3 + (k_1 + k_4 + k_5) \pi \\
		\tilde{\alpha}_4 = (-1)^{k_4} \alpha_4 + (k_2 + k_3 + k_6) \pi \\
		\tilde{\alpha}_5 = (-1)^{k_5} \alpha_5 + (k_3 + r_2) \pi \\
		\tilde{\alpha}_6 = (-1)^{k_6} \alpha_6 + (k_4 + r_3) \pi 
	\end{dcases*}
\end{equation}
where $\textbf{ r} = \{r_i\}_{i=1}^3$ and can be used to mask the real outcomes of the computation. Note that finding these relations, and thus identifying MBQC-related sampling problems, is computationally efficient because of the graph structure of the problem.

For example, we select the original angle set---randomly generated---to be  $\alpha = \{\frac{3}{4}\pi, \frac{7}{3} \pi, \frac{\pi}{3}, 0, \frac{2}{3}\pi, \pi \}$, and the random strings to be $\textbf{k} = \{ 1,0,0,0,1,0\}$, $\textbf{r} = \{ 0,1,1 \}$. Using the relations above we get $\tilde{\alpha} = \{ \frac{5}{4}\pi, \frac{7}{3}\pi, \frac{7}{3}\pi, 0, \frac{\pi}{3},0 \}$. If we simulate the two circuits we obtain:
\begin{align}
\text{Pr}(b_5, b_6) &=
	 \begin{dcases*}
		0.207 & \text{if $b_5, b_6 = (0,0)$}, \\
		0.393 & \text{if $b_5, b_6 = (0,1)$}, \\
		0.043 & \text{if $b_5, b_6 = (1,0)$}, \\
		0.357 & \text{if $b_5, b_6 = (1,1)$},
	\end{dcases*}
\end{align}
and
\begin{align}
\text{Pr}(b_6, b_2, b_5) &=
	 \begin{dcases*}
		0.179 & \text{if $b_2, b_5, b_6 = (0,0,0)$}, \\
		0.021 & \text{if $b_2, b_5, b_6 = (0,0,1)$}, \\
		0.196 & \text{if $b_2, b_5, b_6 = (0,1,0)$}, \\
		0.104 & \text{if $b_2, b_5, b_6 = (0,1,1)$}, \\
		0.060 & \text{if $b_2, b_5, b_6 = (1,0,0)$}, \\
		0.064 & \text{if $b_2, b_5, b_6 = (1,0,1)$}, \\
		0.065 & \text{if $b_2, b_5, b_6 = (1,1,0)$}, \\
		0.311 & \text{if $b_2, b_5, b_6 = (1,1,1)$}.
	\end{dcases*}
\end{align}
The strings we have to compare are the following 
\begin{align}
\text{Pr}(0,0)_{\mathcal{C}_{\text{a}}} &= 2\cdot \text{Pr}(0 \oplus r_1,0 \oplus r_2 ,0 \oplus r_3)_{\mathcal{C}_{\text{b}}} \nonumber \\ 
&= 2 \cdot\text{Pr}(0,1,1)_{\mathcal{C}_{\text{b}}} ,\\
\text{Pr}(0,1)_{\mathcal{C}_{\text{a}}} &= 2 \cdot\text{Pr}(0 \oplus r_1,0 \oplus r_2 ,1 \oplus r_3)_{\mathcal{C}_{\text{b}}} \nonumber\\ 
&= 2 \cdot\text{Pr}(0,1,0)_{\mathcal{C}_{\text{b}}} ,\\
\text{Pr}(1,0)_{\mathcal{C}_{\text{a}}} &= 2\cdot \text{Pr}(0 \oplus r_1,1 \oplus r_2 ,0 \oplus r_3)_{\mathcal{C}_{\text{b}}} \nonumber\\ 
&= 2\cdot \text{Pr}(0,0,1)_{\mathcal{C}_{\text{b}}} ,\\
\text{Pr}(1,1)_{\mathcal{C}_{\text{a}}} &= 2\cdot \text{Pr}(0 \oplus r_1,1 \oplus r_2,1 \oplus r_3)_{\mathcal{C}_{\text{b}}} \nonumber \\
&= 2\cdot \text{Pr}(0,0,0)_{\mathcal{C}_{\text{b}}} .
\end{align}
By checking the outcomes above, we confirm the correctness of the relations.

\section{The squared $\ell^2$ distance}
The central figure of merit for scalably comparing MBQC-related circuits is the squared $\ell^2$-distance introduced in the main text Eq.~\eqref{eq:norm}. In general, the number of output qubits $n_{O_1}$ and $n_{O_2}$ for the two circuits will differ and a subset of $n_c$ of these qubits will be in the output set of both computations. For example the circuits in Fig.~\ref{Fig:SuppBox} have $n_{O_a}=2$ and $n_{O_b}=5$ output qubits, respectively, and $n_{c}=1$ qubit that is in both output sets. There are therefore $n_v=n_{O_a}+n_{O_b}-n_{c}=6$ qubits (in the underlying MBQC) that must be considered for estimating the squared $\ell^2$-distance in Eq.~\eqref{eq:norm} in the main text. Taking the expectation value we can estimate the three resulting terms independently.

A crucial observation is that, due to a generalization of the birthday paradox~\cite{birthday}, the term $\vec{p}_j\cdot\vec{p}_j$ is related to the probability of obtaining the same output string from the $j^\text{th}$ device twice. Note that from the experiment we obtain strings of length $n_{O_j} < n_v$. Using the fact that in the MBQC picture the outcomes of non-output qubits are uniformly random, the probability of a collision between the experimental strings of length $n_{O_j}$ can trivially be related to the probability of a collision between the strings of length $n_v$ in the underlying MBQC. This requires a number of runs that scales as $O(2^{n_{O_j}/2})$.

Estimating the term $\vec{p}_1\cdot\vec{p}_2$, on the other hand, requires us to consider collisions among the strings of length $n_v$. This is achieved by randomly fixing the values of non-output qubits for either circuit (which corresponds to running different computations due to MBQC-corrections for non-zero outcomes) for each sampling run. In the example of  Fig.~\ref{Fig:SuppBox}, circuit $a$ requires randomly fixing 4 of the 6 output qubits and measuring the others, while circuit $b$ requires fixing 1 of the 6 output qubits and measuring the rest. Using this technique, one can estimate the probability for a collision among the $2^6$ possible strings $m$ in the underlying MBQC between the two devices with a number of runs that scales as $O(2^{(n_{O_1}+n_{O_2}-n_c)/2})$.

\section{Sub-sampling}
\begin{figure}[h]
\begin{center}
\includegraphics[width=\columnwidth]{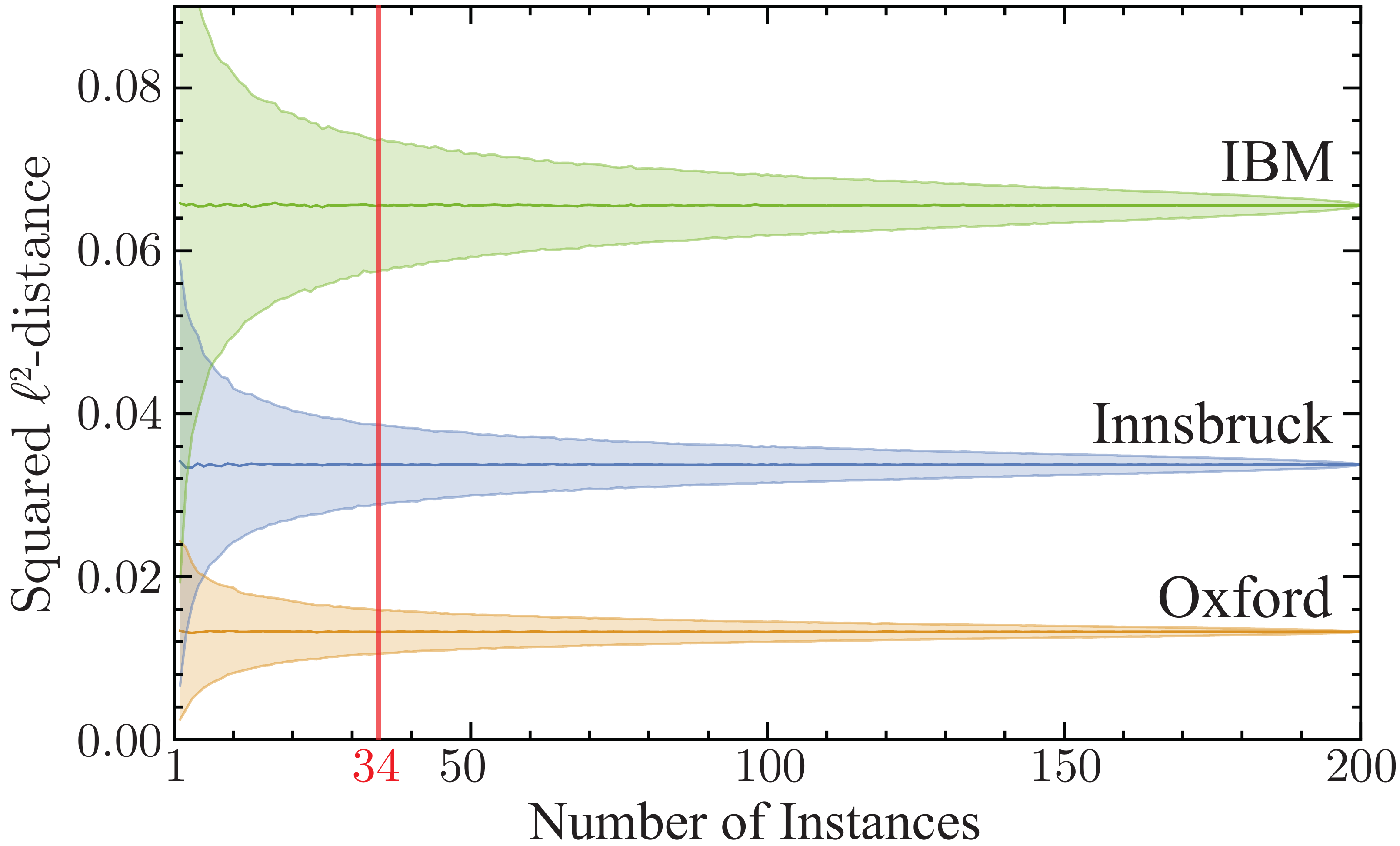}
\caption{Averaged squared $\ell^2$-distances for different numbers of averaged instances. Shown are data for the Innsbruck self-verification (blue), the cross-verification between Innsbruck and Oxford (orange), and the cross-verification between Innsbruck and IBM (green). The solid lines correspond to the mean values over all samples of a fixed size and the shaded regions depict the one-sigma spread of these values. Statistical uncertainties are not taken into account in this analysis.}
\label{fig:subsampling}
\end{center}
\end{figure}

The data presented in the main text was obtained by averaging the squared $\ell^2$-distances obtained from 34 out of 200 randomly chosen instances over the related circuits. Averaging over more instances would increase the confidence in the final estimate, however, comes with an additional resource overhead. To investigate this trade-off, we estimate the squared $\ell^2$-distances from subsets of varying sizes up to the full dataset. The results, shown in Fig.~\ref{fig:subsampling}, indicate quick convergence to the mean value over the full dataset. In particular, the choice of 34 instances is sufficient to clearly distinguish the different devices.

\section{Complementary results}
Here we report data related to the evaluation of quantum circuits equivalent to closed lattice cluster states, without performing the respective measurement based quantum computation. Specifically we consider the closed 2D cluster states involving 8 and 10 qubits shown in Fig.~\ref{Fig:SuppBox}. We refer to those as Box Cluster $2$x$4$ and Box Cluster $2$x$5$, respectively, with $2$x$j$ ($j=[4,5]$) labels the height and width of the cluster. 

Different types of circuits are performed on pairs of quantum devices. In the following table all the MBQC-related devices with the implemented circuits specifications (input qubits and circuit depth) are reported. 

\begin{center}
\begin{tabular}{>{\centering\arraybackslash}p{2.2cm}|>{\centering\arraybackslash}p{1.2cm} >{\centering\arraybackslash}p{1.2cm} >{\centering\arraybackslash}p{2cm}}
Box Cluster &  input & depth & Q. device \\
\hline
\multirow{4}{*}{$2$x$4$} &2 & 4  & Oxford \\
&2 & 4  & Innsbruck \\
 &4 & 2  & IBM \\
\hline
\multirow{2}{*}{$2$x$5$} &2 & 5  & Oxford \\
 &5 & 2  & IBM$_2$ \\
\hline
\end{tabular}
\end{center}
Note that the $5$x$2$ cluster was measured using a different IBM device, namely the \emph{ibmqx3}~\cite{ibmqx3} (\emph{IBM Q 16 Rueschlikon}), which we refer to as IBM$_2$. This device has similar specifications as the first IBM quantum processor used here, but allows for computations with up to 16 qubits.

\subsection{Box Cluster}
\begin{figure*}[t!]
\begin{center}
\includegraphics[width=\linewidth]{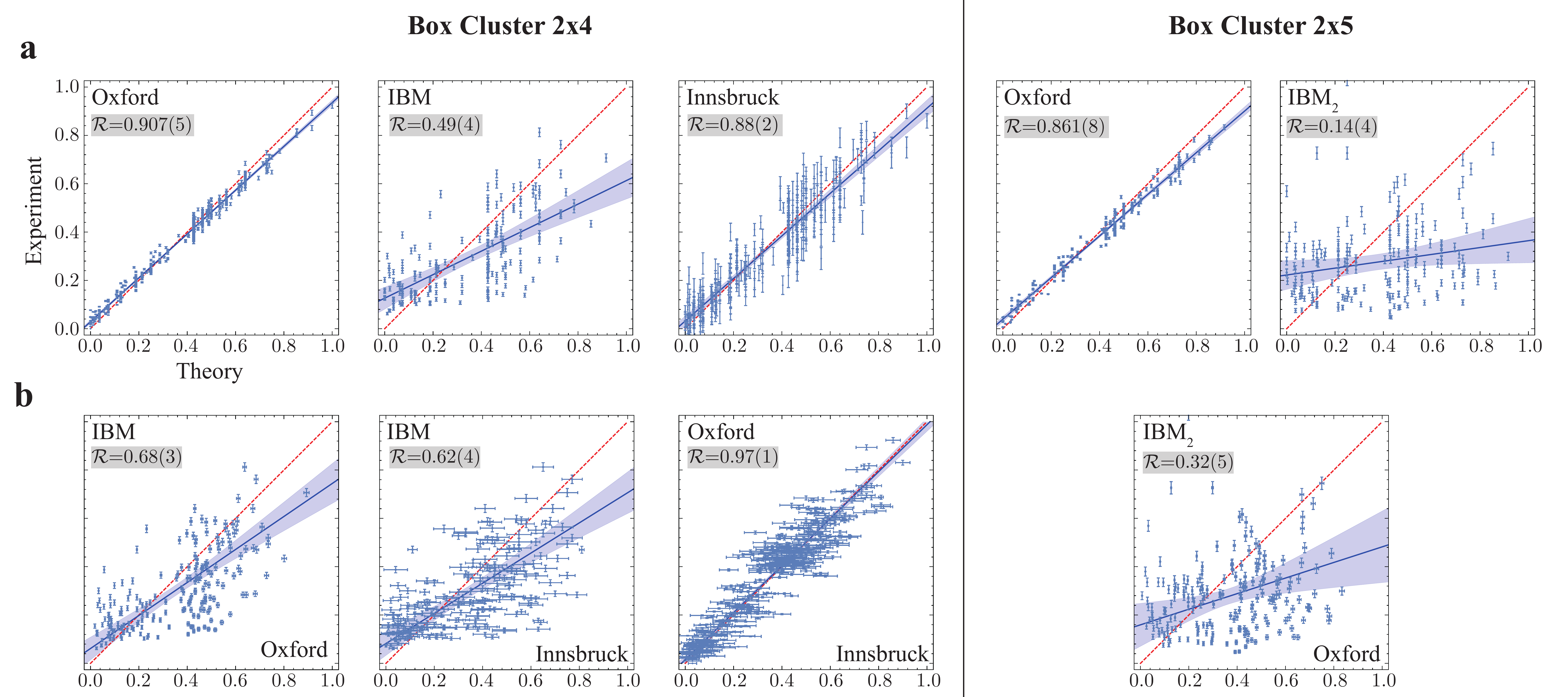}
\caption{Data for cross-verification using the Box Cluster $2$x$4$ (left) and $2$x$5$ (right). \textbf{(a)} Scatterplots of outcome probabilities compared to the theoretical expectation for (from left to right): Oxford, Innsbruck, IBM for the $2$x$4$-cluster, and Oxford, IBM$_2$ for the $2$x$5$ cluster. The squared $\ell^2$-distances $\|\vec{p}_1-\vec{p}_2\|^2$ for these theory comparisons are $0.00200(3)$, $0.0578(5)$, $0.0123(6)$, $0.00400(8)$, $0.145(1)$. This is in good agreement with the trends seen from a linear least-squares regression (blue line) quantifying the deviation from the ideal correlation (red dashed line), with the resulting regression slopes $\mathcal{R}$ with 1-sigma uncertainties given in the top left corner of the respective figure panel. \textbf{(b)} Scatterplots of the outcomes probabilities for two-by-two cross-check verification between Oxford--IBM, Innsbruck--IBM, and Innsbruck--Oxford for the $2$x$4$ cluster, and Oxford--IBM$_2$ for the $2$x$5$ cluster. The squared $\ell^2$-distances $\|\vec{p}_1-\vec{p}_2\|^2$ for the cross-validations are $0.0504(6)$, $0.060(2)$, $0.0115(6)$, $0.118(1)$. This is in good agreement with the regression coefficients $\mathcal{R}$ obtained from linear total least squares regression (blue line) given in the top left corner of the respective figure panel. Experimental error bars correspond to 1-sigma statistical uncertainty and the blue shaded bands represent 3-sigma mean prediction intervals for the regression.}
\label{fig:cluster2x4}
\end{center}
\end{figure*}

As in the main text, the equivalences for the outcome probabilities obtained from the two circuits based on the Box Cluster $2$x$4$ are: 
\begin{align*}
\text{Pr}(0,0)_{\mathcal{C}_{2x4}} = 4 \cdot \text{Pr}(0,0,0,0)_{\mathcal{C}_{4x2}},\\
\text{Pr}(0,1)_{\mathcal{C}_{2x4}} = 4 \cdot \text{Pr}(0,0,0,1)_{\mathcal{C}_{4x2}}. 
\end{align*}
Similarly, for the Box Cluster $2$x$5$ we obtain: 
\begin{align*}
\text{Pr}(0,0)_{\mathcal{C}_{2x5}} = 8 \cdot \text{Pr}(0,0,0,0,0)_{\mathcal{C}_{5x2}},\\
\text{Pr}(0,1)_{\mathcal{C}_{2x5}} = 8 \cdot \text{Pr}(0,0,0,0,1)_{\mathcal{C}_{5x2}}. 
\end{align*}
We ran 100 $\mathcal{C}_{2x4}$ circuits on the Oxford and Innsbruck machines, with the 100 MBQC-related $\mathcal{C}_{4x2}$ circuits run on the IBM processor. For the $2$x$5$ case we ran 100 $\mathcal{C}_{2x5}$ circuits on the Oxford machine and the 100 MBQC-related $\mathcal{C}_{5x2}$ circuits on the IBM processor. In each case, pair-wise cross-check verification was performed between all devices, as well as individual comparisons to theory. Scatterplots of the outcome probabilities are shown in Fig.~\ref{fig:cluster2x4} and all relevant numerical values are given in the caption of that figure. As in the main text, we find that the squared $\ell^2$-distance provides a very good estimate of the true performance of the devices in agreement with the theory simulation.

\end{document}